\documentclass[a4]{elsart}

\usepackage{graphicx}
\usepackage{subfigure}
\usepackage[latin1]{inputenc}
\usepackage{amsmath}
\usepackage{cite}
\usepackage{xspace}
\usepackage{amssymb}

\newcommand{\nSc}{n_{\mathrm{Sc}}}

\newcommand{\nX}{n_{\mathrm{X}}}
\newcommand{\ie}{{i.e.}\@\xspace}
\newcommand{\cf}{{cf.}\@\xspace}
\newcommand{\Fig}[1]{Fig.~\ref{#1}}
\newcommand{\Eq}[1]{Eq.~\ref{#1}}
\begin{document}

\begin{frontmatter}

\title{Precipitation kinetics of Al$_3$Zr and Al$_3$Sc in aluminum alloys
modeled with cluster dynamics}

\author[SRMP]{Emmanuel Clouet\corauthref{email}},
\author[SRMP]{Alain Barbu},
\author[LTPCM]{Ludovic Laé},
\author[HC]{Georges Martin}

\corauth[email]{emmanuel.clouet@cea.fr}

\address[SRMP]{Service de Recherches de Métallurgie Physique, CEA/Saclay, \\
91191 Gif-sur-Yvette, France}
\address[LTPCM]{LTPCM/ENSEEG, INPG, B.P. 75, \\
38402 Saint Martin d'Hères, France}
\address[HC]{Cabinet du Haut-Commissaire, CEA/Siège,  \\
31-33, rue de la Fédération, \\
75752 Paris cedex 15, France}

\begin{abstract}
Precipitation kinetics of Al$_3$Zr and Al$_3$Sc in aluminum 
supersaturated solid solutions is studied using cluster dynamics,
a mesoscopic modeling technique which describes 
the various stages of homogeneous precipitation
by a single set of rate equations.
The only parameters needed are the interface free energy
and the diffusion coefficient which are deduced from an atomic model
previously developed to study the same alloys.
A comparison with kinetic Monte Carlo simulations 
based on the vacancy diffusion mechanism
shows that cluster dynamics correctly predicts the precipitation kinetics
provided a size dependent interface free energy is used.
It also manages to reproduce reasonably well existing experimental data.
\end{abstract}

\begin{keyword}
precipitation \sep kinetics \sep aluminum alloys \sep cluster dynamics
\PACS
64.60.Cn 
\sep 64.60.-i 
\sep 64.70.Kb 
\sep 64.75.+g 
\end{keyword}
\end{frontmatter}

\section{Introduction}

Transition elements are added 
to aluminum alloys so as to obtain small ordered precipitates and by this way 
decrease the alloy sensitivity to recrystallization. 
Among these elements, zirconium and scandium are 
the most efficient ones. They lead to the precipitation 
of respectively Al$_3$Zr and Al$_3$Sc with the L1$_2$ structure 
\cite{ROB01,RYU69,NES72,HYL92,MAR01,NOV01}.
As recrystallization is controlled by the precipitate density and size,
it is important to model precipitation kinetics in Al-Zr and Al-Sc alloys,
in order to optimize the heat treatment and better control the alloy final state.

In a previous study \cite{CLO04,CLO04T}, one of us built an atomic model 
for the above two binary alloys and studied  precipitation kinetics
using Monte Carlo simulations based on the vacancy diffusion mechanism. 
The drawback of such an approach is that only short annealing times 
can be simulated which implies using high supersaturations.
We showed that the classical nucleation theory manages 
to reproduce atomic simulations giving thus good confidence in the 
ability of mesoscopic models to quantitatively predict
the nucleation stage of precipitation kinetics.
In the present work, starting from the same atomic model,
we build a mesoscopic modeling, the cluster dynamics \cite{GOL95,MAT97,BAR04}
which is based on a set of rate equations describing in the same framework
the three different stages of precipitation, \ie nucleation, growth and coarsening.
It requires only a limited number of parameters, the interface free energy
and the solute diffusion coefficients, which can be quite easily deduced 
from atomic parameters.
The range of supersaturations and annealing times that can be simulated 
are thus extended allowing a comparison with experimental data. 
Cluster dynamics technique has already been used to model 
precipitation kinetics in Al-Zr and Al-Sc alloys 
in the spirit of nucleation theory \cite{LAE04}.
The way we use cluster dynamics in the present paper is distinct, 
as will be shown below, and is based on the description of the alloy
as a gas of clusters. Unlike the nucleation theory, our approach
leaves no arbitrary choice for the parameters used.

In the first part of this article, cluster dynamics modeling is presented. 
We show then how precipitation kinetics obtained with this method
compare with kinetic Monte Carlo simulations for both Al-Zr and Al-Sc systems
and try to improve the kinetic description so as to obtain a better agreement. 
Finally, available experimental data are compared with simulations.

\section{Cluster dynamics}

In its strict sense, cluster dynamics rests on the description of the alloy
undergoing phase separation as a gas of solute clusters which exchange
solute atoms by single atom diffusion\cite{GOL95,MAT97,BAR04}. 
Clusters are assumed to be spherical and 
are described by a single parameter, their size or the number 
$\nX$ of solute atoms $\mathrm{X}$ they contain.
In such a description, there is no precisely defined distinction
between the solid solution on the one hand and the precipitates on the other hand,
at variance with the classical nucleation theory: the distribution of cluster sizes
is the only quantity of interest (for a detailed discussion see Ref. \citen{MAR04}).
In an undersaturated solid solution of nominal concentration $x^0_{\mathrm{X}}$
at equilibrium, the cluster size distribution is given by (\cf appendix \ref{distri_app})
\begin{equation}
\bar{C}_{\nX}(x^{0}_{\mathrm{X}}) = 
\exp{\left[-\left( G_{\nX} - 2 \nX \mu(x^{0}_{\mathrm{X}}) \right) / kT\right]},
\label{distri_amas_eq}
\end{equation}
$G_{\nX}$ being the free energy of a cluster containing $\nX$ atoms 
and $\mu(x^{0}_{\mathrm{X}}) = \left( \mu_{\mathrm{X}}(x^{0}_{\mathrm{X}})
-\mu_{\mathrm{Al}}(x^{0}_{\mathrm{X}}) \right)/2$
the effective chemical potential in the 
solid solution, \ie half the difference between the chemical potentials
of the solute and the solvent atoms.
In a supersaturated system, the cluster size distribution evolves in time 
as discussed below.

\subsection{Master equation}

The cluster dynamics technique describes the precipitation kinetics
thanks to a master equation giving the time evolution 
of the cluster size distribution\cite{GOL95,MAT97,BAR04}. 
When only monomers can migrate, which is the case for Al-Zr
as well as Al-Sc systems \cite{CLO04}, the probability $C_{\nX}$ 
to observe a cluster containing $\nX$ solute atoms obeys
the differential equations
\begin{subequations}
\begin{align}
\frac{\d C_{\nX}}{\d t} &= J_{\nX-1\to\nX} - J_{\nX\to\nX+1} ,\quad \forall\ \nX\geq2 \\
\frac{\d C_{1}}{\d t} &= - 2 J_{1\to2} - \sum_{\nX\geq2}{J_{\nX\to\nX+1}},
\end{align}
\label{eq:DA}
\end{subequations}
where $J_{{\nX} \to {\nX}+1}$ is the cluster flux from the class of size $\nX$ 
to the class $\nX+1$.
This flux can be written
\begin{equation}
J_{{\nX} \to {\nX}+1} = \beta_{\nX} C_{\nX} - \alpha_{{\nX}+1} C_{{\nX}+1}
\label{eq:flux_DA} ,
\end{equation}
$\beta_{\nX}$ being the probability per unit time for one solute atom 
to impinge on a cluster of size $\nX$
and $\alpha_{\nX}$ for one atom to leave a cluster of size $\nX$.

In order to handle large cluster sizes, the set of differential equations 
\ref{eq:DA} and \ref{eq:flux_DA} are integrated by considering 
cluster sizes in a discrete way up to a maximal size ($n_{\mathrm{X}}\sim100$)
and by using a continuous variable with an increasing step beyond \cite{HAR02b}.
\subsection{Condensation rate}

When the solute long-range diffusion controls the precipitation kinetics,
the condensation rate is obtained by solving the diffusion problem in the solid 
solution around a spherical precipitate (\cf appendix \ref{condensation_app}).
For a cluster of radius $r_{\nX}$, this condensation rate takes the form \cite{WAI58,MAR78}
\begin{equation}
\beta_{\nX} = 4 \pi r_{\nX} \frac{D_{\mathrm{X}}}{\Omega} C_1,
\label{eq:beta_n}
\end{equation}
where $D_{\mathrm{X}}$ is the diffusion coefficient of the solute at infinite 
dilution of the solid solution 
and $\Omega$ is the mean atomic volume corresponding to one lattice site.
We checked by studying growth of planar precipitates
that precipitation kinetics can truly be assumed to be controlled
by solute long range diffusion 
and that the diffusion coefficients deduced from kinetic Monte Carlo simulations
are in good agreement with the experimental ones entering equation \ref{eq:beta_n}
\cite{NAS04}.

\subsection{Evaporation rate}

By contrast with the condensation rate, the evaporation rate $\alpha_{\nX}$ cannot
be obtained directly. It has to be deduced from $\beta_{\nX}$ using 
the equilibrium cluster size distribution (Eq.~\ref{distri_amas_eq}). 

The evaporation rate is obtained assuming that  
it is an intrinsic property of the cluster and
therefore does not depend on the solid solution 
surrounding the cluster\footnote{Sometimes, one assumes instead
that a constraint equilibrium exists for the clusters in the supersaturated solid solution
and that the expression \ref{distri_amas_eq} holds even for concentrations higher 
than the solubility limit $x^{eq}_{\mathrm{X}}$.
Katz and Wiedersich \cite{KAT77} pointed that this assumption can
lead to the same expression of the evaporation rate as the one we used.
In particular, this is true when the condensation rate varies linearly
with the monomer concentration which is the case here.}.
This means that the cluster has enough time to explore all its configurations
between the arrival and the departure of a solute atom.
Thus $\alpha_{\nX}$ should not depend on the nominal concentration 
of the solid solution and can be obtained by considering 
any undersaturated solid solution of nominal concentration $x^0_{\mathrm{X}}$.
Such a solid solution is stable. Then there should be no energy dissipation.
This involves that all fluxes $J_{\nX\to\nX+1}$ equal zero.
Using equation \ref{eq:flux_DA}, one obtains
\begin{equation}
\begin{split}
\alpha_{{\nX}+1} &= \bar{\alpha}_{{\nX}+1}(x^{0}_{\mathrm{X}})  \\
&= \bar{\beta}_{\nX}(x^{0}_{\mathrm{X}}) 
\frac{\bar{C}_{\nX}(x^{0}_{\mathrm{X}})}{\bar{C}_{{\nX}+1}(x^{0}_{\mathrm{X}})}, 
\end{split}
\label{an_equilibre_eq}
\end{equation}
where overlined quantities are evaluated in the solid solution at equilibrium.
In particular, the equilibrium cluster size distribution is given by 
equation \ref{distri_amas_eq}.
This finally leads to the following expression of the evaporation rate
\begin{equation}
\alpha_{{\nX}+1} =  4 \pi r_{\nX} \frac{D_{\mathrm{X}}}{\Omega} 
\exp{\left[ (G_{\nX+1} - G_{\nX} - G_{1} ) / kT\right]}.
\label{an_eq}
\end{equation}
As the condensation rate varies linearly with the monomer concentration, 
the contribution of the effective chemical potential
cancels out in the expression \ref{an_equilibre_eq} of $\alpha_{\nX}$:
we recover our starting assumption.
Indeed, all informations concerning the solid solution are contained in 
this chemical potential. 
Therefore $\alpha_{\nX}$ only depends on the clusters free energies 
and not on the overall state of the gas of clusters.
Especially, the evaporation rate is independent of the nucleation free energy,
\ie the free energy gained when nucleating a precipitate out of the solid solution:
such concepts are useful in the classical nucleation theory but play no role in cluster dynamics.
This can be seen by dividing the cluster formation free energy 
into a volume and an interface contributions \cite{CLO04},
\begin{equation}
G_{\nX} - 2 \nX \mu(x^{0}_{\mathrm{X}}) 
= 4\nX \Delta G^{nuc}(x^{0}_{\mathrm{X}}) 
+ \left(36\pi {\nX}^2\right)^{1/3} a^2 \sigma_{n_{\mathrm{X}}} ,
\label{capillary_eq}
\end{equation}
$\Delta G^{nuc}(x^{0}_{\mathrm{X}})$ being the nucleation free energy 
of a solid solution of nominal concentration $x^{0}_{\mathrm{X}}$,
$\sigma_{n_{\mathrm{X}}}$ the interface free energy of a cluster
containing $\nX$ solute atoms and $a$ the lattice parameter.
This leads to the following expression of the evaporation rate
\begin{equation}
\begin{split}
\alpha_{{\nX}+1} =&  4 \pi r_{\nX} \frac{D_{\mathrm{X}}}{\Omega} \\
&\exp{\left[ (36\pi)^{1/3} a^2 \left(
(\nX+1)^{2/3} \sigma_{\nX+1} - {\nX}^{2/3} \sigma_{\nX} - \sigma_{1} \right) / kT\right]}.
\end{split}
\label{an_eq2}
\end{equation}

Looking at the expression \ref{eq:beta_n} of the condensation rate 
and the expression \ref{an_eq2} of the evaporation rate, 
one sees that the only parameters needed by cluster dynamics are the diffusion
coefficient and the interface free energy.
There is no need to know the nucleation free energy\footnote{Some cluster dynamics studies
\cite{LAE04} makes the evaporation rate depend on the nucleation free energy
because a constrained equilibrium of the cluster size distribution has been
considered so as to reproduce the instantaneous monomer concentration $C_1$
and not the total solid solution concentration $x^0_{\mathrm{X}}$ as it should be.} in opposition to 
other mesoscopic models based on classical nucleation theory following
Langer and Schwartz approach \cite{LAN80} as modified by 
Kampmann and Wagner \cite{WAG91,DES99,ROB01,ROB03}.
In cluster dynamics, thermodynamics of the solid solution is described thanks to a lattice 
gas model and therefore the nucleation free energy results from this description 
and is not an input of the modeling.

\subsection{Definition of precipitates}

The master equation which describes cluster dynamics yields the full 
cluster distribution as a function of time.
In order to follow the precipitation kinetics and make the link
with experimental results, it is convenient to define mean values such as
the precipitates mean size or their density so as to characterize kinetics 
by the evolution of these quantities. 
Therefore, the solid solution has to be discriminated from the precipitates:
one needs to define an arbitrary threshold size $\nX^*$ above which clusters represent 
stable precipitates and below which they represent fluctuations in the solid solution. 
One natural choice for this threshold size is the initial critical size given 
by the classical nucleation theory,
\begin{equation}
	\nX^* = -\frac{\pi}{6}\left(\frac{a^2\sigma_{\nX^*}}{\Delta G^{nuc}(x^0_{\mathrm{X}})}\right).
\end{equation}
This requires to calculate for a given temperature 
and a given composition of the solid solution the nucleation free energy.
This cannot be directly done with the lattice gas model used by cluster dynamics 
as the effective chemical potential 
$\mu_{\mathrm{X}}(x^0_{\mathrm{X}})$ appearing in equation \ref{capillary_eq}
and needed to obtain $\Delta G^{nuc}(x^0_{\mathrm{X}})$ is implicit.
This potential can be deduced from the simulation when a stationary
cluster size distribution is observed\footnote{The effective
chemical potential is then linked to the instantaneous monomer concentration
by the relation $\mu = \left[ G_1+kT\ln{(C_1)}\right]/2$.} but it is not 
known \emph{a priori}.
Nevertheless, the nucleation free energy can be estimated by other means 
and in particular thanks to different mean field approximations. 
In reference \citen{CLO04} it was shown that the cluster 
variation method (CVM) \cite{KIK51,SAN78} in the tetrahedron-octahedron approximation leads
to a good prediction of $\Delta G^{nuc}(x^0_{\mathrm{X}})$. 
Therefore the value of the threshold size is chosen equal
to that of the critical size in the classical nucleation theory
when CVM is used to calculate the nucleation free energy.
Thermodynamics, and then the nucleation free energy, deduced from CVM can differ
from the ones really corresponding to cluster dynamics. Nevertheless, 
this nucleation free energy is only used to calculate a 
threshold size and as the aim of this paper is to compare 
precipitation kinetics obtained with Monte Carlo and cluster 
dynamics simulations, this is not a real issue. 
The key point is to use the same threshold size when comparing
the two modeling techniques.

\section{Cluster dynamics simulations}

In references \citen{CLO04} and \citen{CLO04T} one of us developed an atomic kinetic model 
for Al-Zr and Al-Sc systems. Parameters were 
deduced from experimental data (Zr and Sc solubility limits in Al 
and diffusion coefficients) as well as from ab-initio calculations
(Al$_3$Zr and Al$_3$Sc free energy of formation).
So as to compare the precipitation kinetics obtained from 
cluster dynamics with kinetic Monte Carlo simulations we have to
deduce from this atomic model the interface free energies 
between the precipitates and the aluminum matrix.
As for diffusion coefficients, we use in the cluster dynamics simulations
the experimental values known for Zr and Sc impurity diffusion in 
aluminum \cite{LANDOLT26,MAR73,FUJ97}:
\begin{eqnarray}
D_{\mathrm{Zr}^*} &=& 728\times10^{-4}
\exp{(-2.51\mathrm{~eV}/kT)}
\mathrm{~m}^{2}.\mathrm{s}^{-1} \nonumber , \\
D_{\mathrm{Sc}^*} &=& 5.31\times10^{-4}
\exp{(-1.79\mathrm{~eV}/kT)} \mathrm{~m}^{2}.\mathrm{s}^{-1}
\nonumber .
\end{eqnarray}
As shown in Ref.~\citen{CLO04,CLO04T}, the diffusion model used 
in kinetic Monte Carlo does reproduce the above values.

\subsection{Interface free energy}

The interface free energy can be deduced from the atomic model 
by computing the free energies
corresponding to planar interfaces in the three most dense packing
orientations \{111\}, \{110\} and \{100\} within the Bragg-Williams approximation 
and then by using the Wulff construction to obtain an isotropic average interface free energy 
$\bar{\sigma}$ \cite{CLO04}.
Such an interface free energy depends on the temperature for two reasons:
atomic interactions used to obtain $\bar{\sigma}$ vary with temperature
and there is a configurational entropy contribution due to the relaxation 
of the interface.
On the other hand it does not depend on the size of the cluster
and corresponds to the asymptotic limit of $\sigma_{\nX}$.
A direct calculation considering the partition functions of small clusters shows that 
$\sigma_{\nX}$ slightly deviates from this asymptotic value at small sizes \cite{CLO04}.
In the following, we test effects on cluster dynamics results due to this slight dependence
of the interface free energy with the cluster size.
Some authors \cite{GOL95,GOL00,BAR04} already used an interface free energy
depending on the cluster size so as to model copper precipitation in iron by cluster dynamics.
In these studies, the size dependence was introduced to take into 
account the change with size of the precipitate structure.
This is not the purpose here as precipitates keep their L1$_2$ structure 
whatever their size. We want to show how sensitive are cluster dynamics
simulations to the interface free energy and why one needs to go 
beyond a model using a constant parameter.

\begin{figure}[!bp]
\begin{center}
\includegraphics[width=0.8\linewidth]{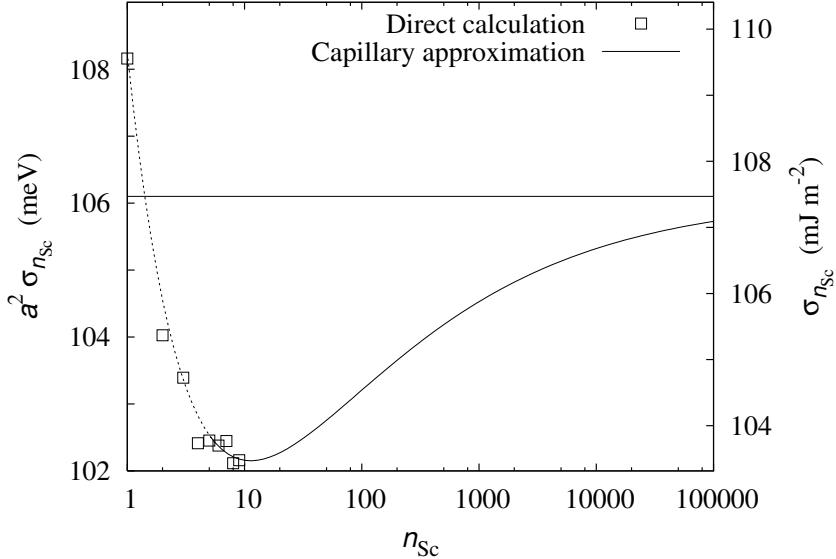}
\end{center}
\caption{Variation with the cluster size $\nSc$ of the interface
free energy between the solid solution and Al$_3$Sc at $T=500$°C. 
Symbols correspond to $\sigma_{\nSc}$ as given by the direct calculation
of the cluster formation free energy (Eq.~20 of Ref.~\citen{CLO04})
and lines to the capillary approximation (Eq.~\ref{capillary_sigma_eq})
as well as to its asymptotic limit.}
\label{sigma_fig}
\end{figure}

In this purpose, for clusters containing no more than 9 solute atoms, 
the interface free energy is computed using Eq.~20 of 
Ref.~\citen{CLO04}.
For clusters of size $\nX\geq10$, the interface free energy is obtained using
an extension of the capillary approximation,
\begin{equation}
\sigma_{\nX} = \bar{\sigma}\left( 1 + c\ \nX^{-1/3} + d\ \nX^{-2/3} \right),
\label{capillary_sigma_eq}
\end{equation}
where $c$ and $d$ respectively correspond to the \emph{line} and \emph{point} contributions 
\cite{PER84}.
The asymptotic value $\bar{\sigma}$ used is the one previously calculated 
thanks to the Bragg-Williams approximation and the Wulff construction \cite{CLO04}
whereas coefficients $c$ and $d$ are obtained by a least square fit of the expression
20 of Ref.~\citen{CLO04} for sizes $5\geq\nX\geq9$.
The curvature correction $c$ obtained with such a procedure is found to be negative
in agreement with classical models \cite{TOL49,BUF55}.
Instead of using such an approximate procedure to obtain 
the interface free energy, this can be done more precisely by sampling thermodynamic
averages with a Monte Carlo algorithm so as to compute the free energy difference
between a cluster of size $n$ and one of size $n+1$ at a given temperature \cite{PER84}. 
Interface free energies obtained by both methods are in good agreement\cite{LEP04}.

As seen in figure \ref{sigma_fig}, the difference between this interface free energy
$\sigma_{\nX}$ depending on the cluster size and its asymptotic limit $\bar{\sigma}$ is 
small. For instance, for the interface between Al$_3$Sc and the solid solution at $T=500$°C, 
the maximal difference is obtained for sizes $\nX\sim10$ and is less than 5\%.

\begin{figure}[!tp]
\begin{center}
	\subfigure[$x^0_{\mathrm{Sc}}=0.4$~at.\% ($n^*_{\mathrm{Sc}}=16$)]
		{\includegraphics[width=0.49\linewidth]{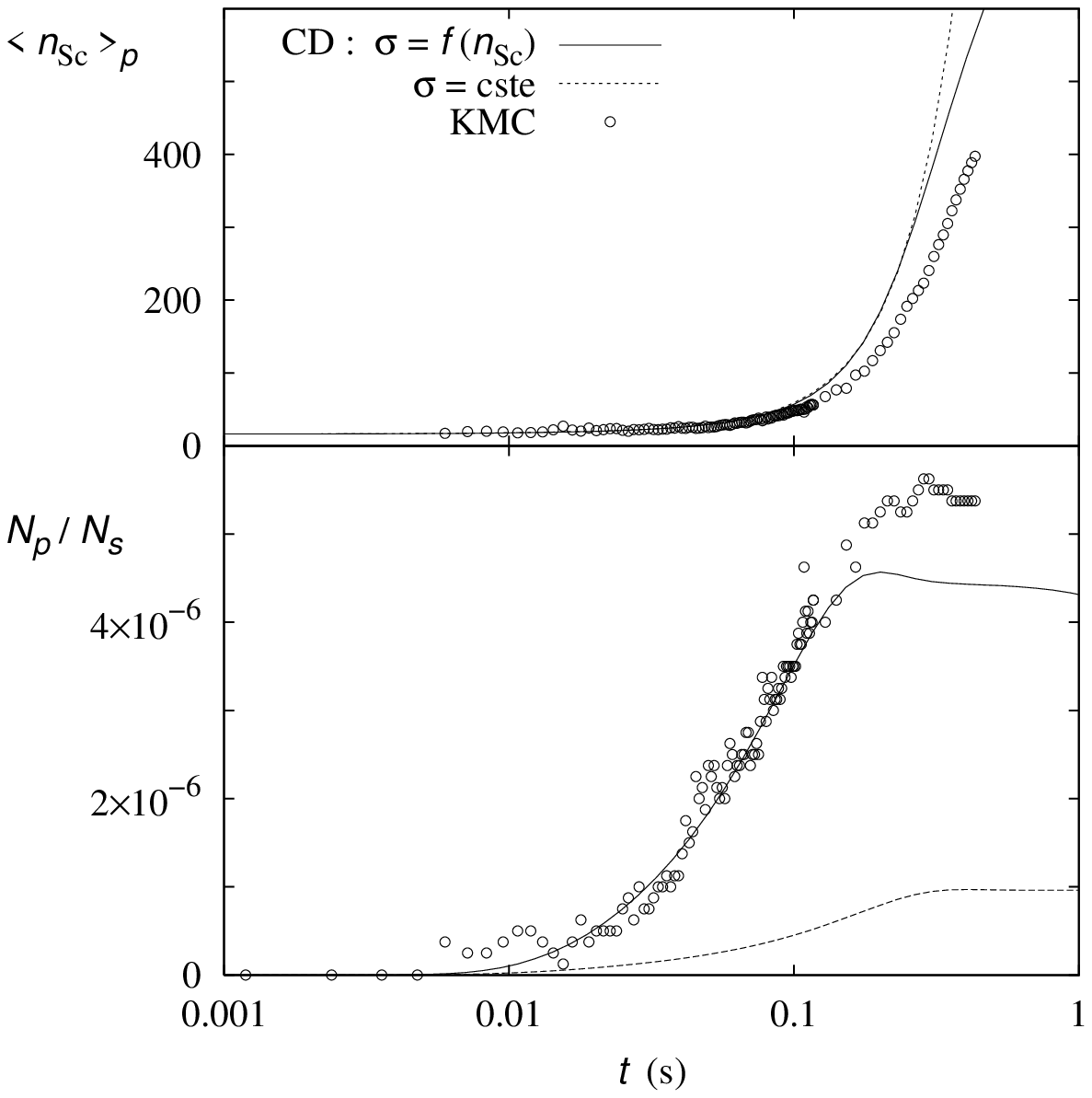}}
	\subfigure[$x^0_{\mathrm{Sc}}=0.75$~at.\% ($n^*_{\mathrm{Sc}}=9$)]
		{\includegraphics[width=0.49\linewidth]{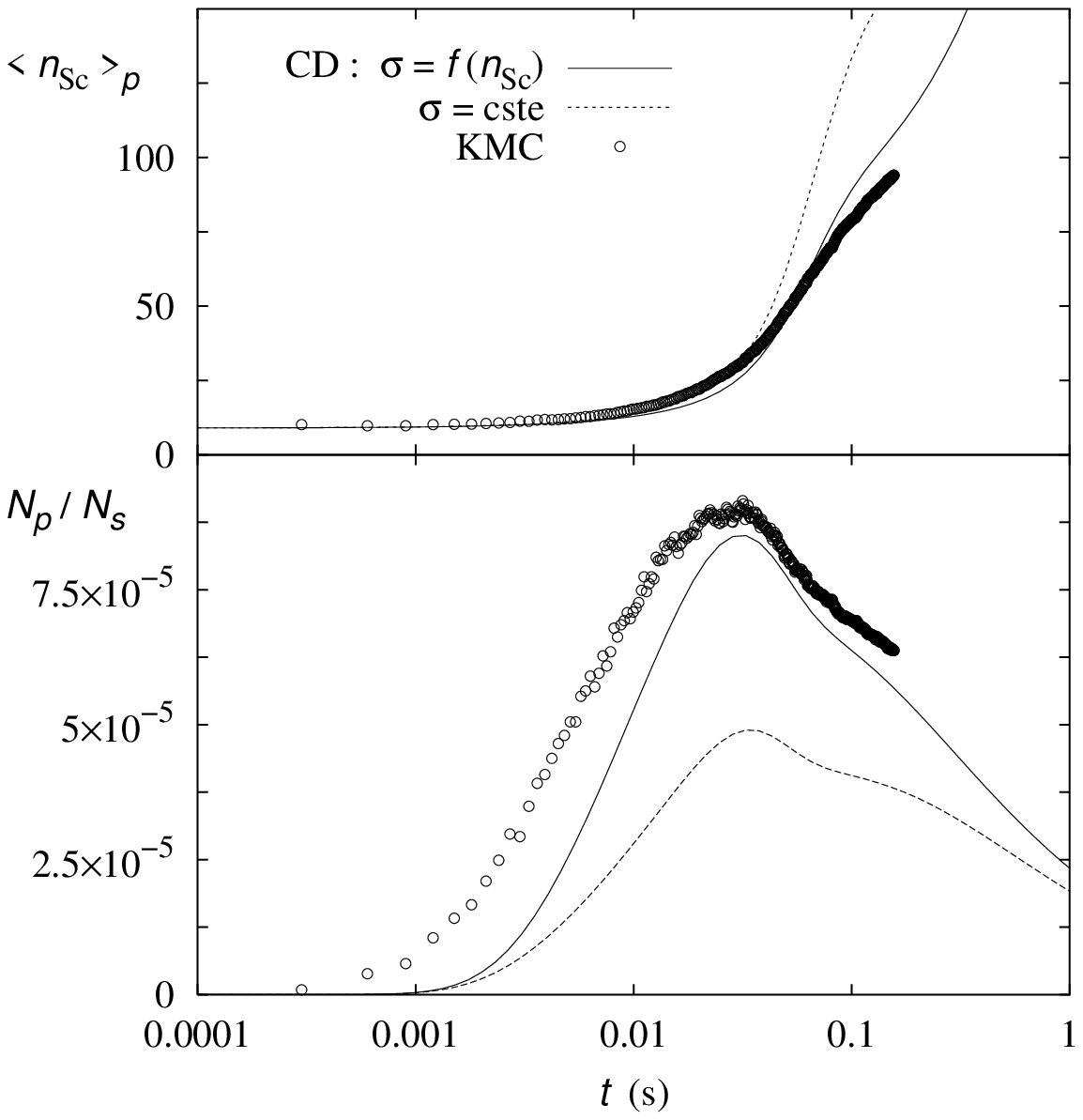}}
\end{center}
\caption{Precipitation kinetics of two supersaturated solid solutions
of nominal concentration $x^0_{\mathrm{Sc}}=0.4$~at.\% and $x^0_{\mathrm{Sc}}=0.75$~at.\%
at $T=500$°C:
evolution with time of the number $N_p$ of precipitates in the simulation
box (normalized by the number of lattice sites $N_s$)
and of their average size $\left<\nSc\right>_p$.
Full lines correspond to cluster dynamics simulations with an interface free energy
depending on the cluster size (Eq.~20 of Ref.~\citen{CLO04}
and \ref{capillary_sigma_eq}),
dotted lines to simulations with a constant interface free energy
and symbols to kinetic Monte Carlo simulations}.
\label{kinetics_sigma_fig}
\end{figure}

We run some cluster dynamics simulations with the interface free energy depending 
on the cluster size as given by Eq.~20 of Ref.~\citen{CLO04}
and Eq. \ref{capillary_sigma_eq} and some other simulations with a constant 
interface free energy taken equal to the asymptotic limit $\bar{\sigma}$.
The comparison with kinetic Monte Carlo simulations (Fig.~\ref{kinetics_sigma_fig})
shows that one obtains a better agreement when taking into account the size dependence: 
the number of precipitates as well as their mean size predicted by both simulation methods 
are then quite close.
Although the size dependence of the interface free energy is small, 
it definitely improves the ability of cluster dynamics to well 
reproduce kinetic Monte Carlo results. 
This is true mainly for the precipitate density, the effect on precipitate mean size
being less pronounced, especially when the supersaturation is low.
The density is more sensitive than the mean size to the distribution of small clusters.
As these clusters are the most affected by the variation with size of the interface energy, 
this explains why the greatest effect observed is for the precipitate density.

In the following, all cluster dynamics simulations will use
this size dependent interface free energy as input parameter.

\subsection{Precipitation kinetics}

\begin{figure}[!bp]
\begin{center}
\subfigure{\includegraphics[width=0.49\linewidth]{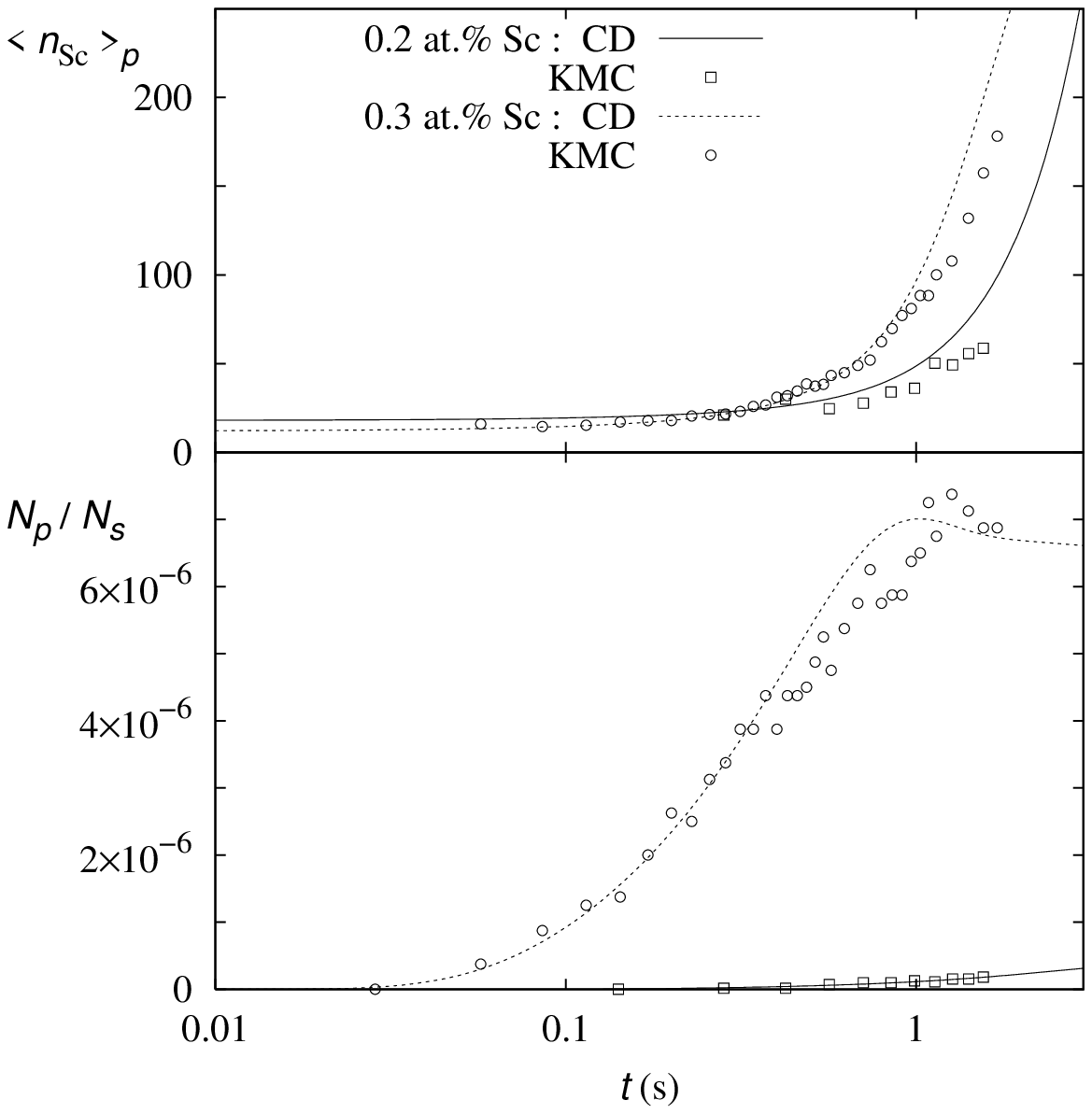}}
\hfill
\subfigure{\includegraphics[width=0.49\linewidth]{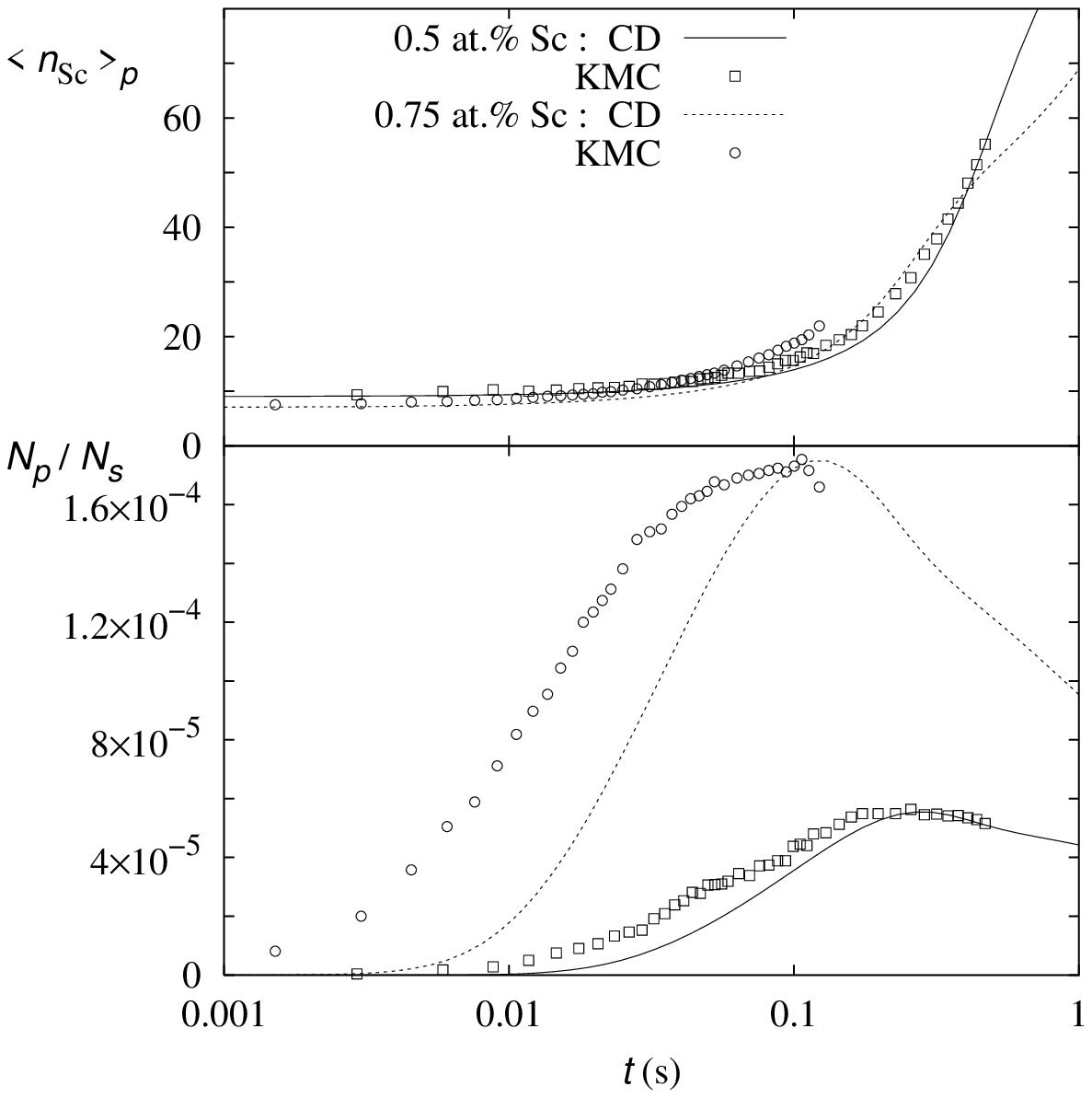}}
\hfill
\subfigure{\includegraphics[width=0.49\linewidth]{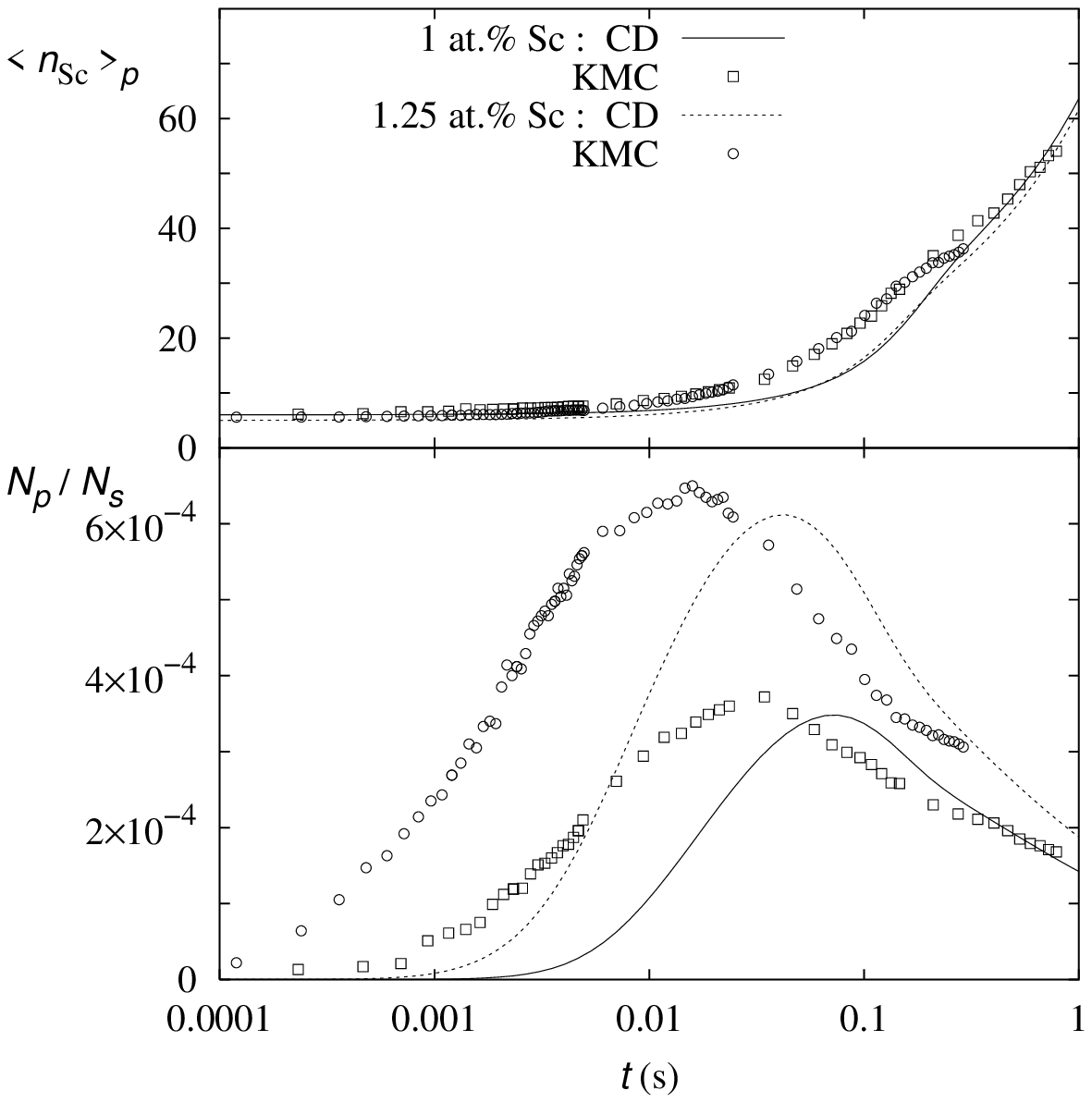}}
\hfill
\end{center}
\caption{Precipitation kinetics for different nominal concentrations 
of a supersaturated aluminum solid solution at $T=450$°C.
Lines correspond to cluster dynamics simulations and symbols to 
kinetic Monte Carlo simulations.}
\label{kinetics_fig1}
\end{figure}

Comparing, at a given temperature, the precipitation kinetics obtained with Monte Carlo
and cluster dynamics simulations for different nominal concentrations of the solid solution 
(\Fig{kinetics_fig1}), one sees that cluster dynamics manages to reproduce the 
variations of the precipitate density as well as the variations of their mean size.
For low supersaturations, the agreement is really good whereas for higher supersaturations
there is a delay: cluster dynamics is too slow compared to kinetic Monte Carlo simulations
by a factor of roughly 2.
Nevertheless, in all cases, the prediction of the precipitate maximal density 
at the transition between the growth and coarsening stages is correct.
Cluster dynamics technique manages to reproduce
the variations over three orders of magnitude of this maximal density
in the range of concentrations considered
in Al-Sc solutions at $T=450$°C (\Fig{kinetics_fig1}).

\begin{figure}[!bp]
\begin{center}
\subfigure[$x^0_{\mathrm{Zr}}=1$~at.\%]{\includegraphics[width=0.49\linewidth]{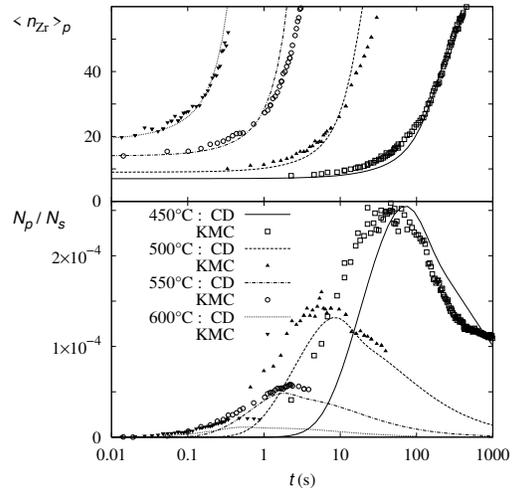}}
\hfill
\subfigure[$x^0_{\mathrm{Zr}}=1.25$~at.\%]{\includegraphics[width=0.49\linewidth]{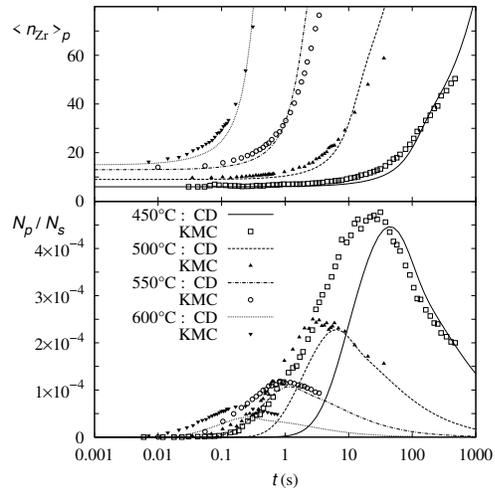}}
\end{center}
\caption{Precipitation kinetics for different temperatures 
of a supersaturated solid solution of nominal concentration
$x^0_{\mathrm{Zr}}=1.$~at.\% and
$x^0_{\mathrm{Zr}}=1.25$~at.\%}
\label{kinetics_fig2}
\end{figure}

Fixing now the nominal concentration of the solid solution and varying 
the temperature, the comparison of results obtained with both simulation techniques
leads to the same conclusions (\Fig{kinetics_fig2}). 
Cluster dynamics results still appear to be too slow at high supersaturations,
but except this time delay, variations of precipitate density and of their mean size are well reproduced.

\clearpage
\subsection{Cluster size distribution}

\begin{figure}[!bp]
\begin{center}
\includegraphics[width=0.8\linewidth]{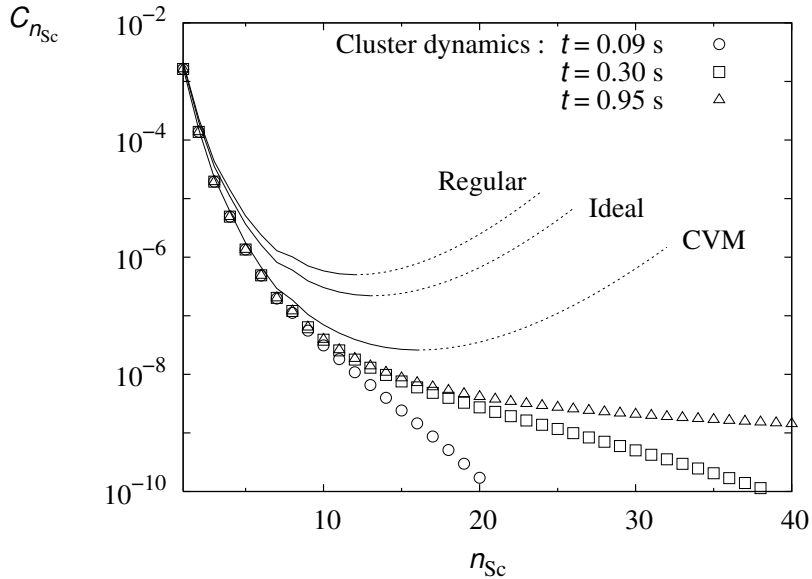}
\end{center}
\caption{Cluster size distribution of an aluminum solid solution of nominal
concentration $x^0_{\mathrm{Sc}}=0.2$~at.\% at $T=450$°C. 
Symbols correspond to cluster dynamics simulations and lines
to prediction of classical nucleation theory with different
mean-field approximations of the nucleation free energy.}
\label{size_distrib_fig}
\end{figure}

Cluster dynamics lead to a stationary cluster size distribution in the solid solution,
\ie for clusters smaller than a critical size $\nX^*$ (\Fig{size_distrib_fig}),
as predicted by classical nucleation theory.
This stationary distribution corresponds to 
a constraint equilibrium existing for sub-critical clusters
but nothing has been supposed concerning clusters bigger than the critical one.
Using the capillary approximation to estimate the cluster formation 
free energy (\Eq{capillary_eq}), classical nucleation theory 
assumes this stationary distribution to be given by equation 
\ref{distri_amas_eq}.
Obviously, in this expression, the interface free energy has to be taken equal to the one
used for cluster dynamics simulations and therefore we consider the size dependent
interface free energy given by Eq.~20 of Ref.~\citen{CLO04}
and Eq.\ref{capillary_sigma_eq}.
As for the nucleation free energy $\Delta G^{nuc}(x^0_{\mathrm{X}})$, 
we already pointed that several thermodynamic approximations can be used to estimate it.
One usually considers ideal or regular solid solution models to calculate 
$\Delta G^{nuc}(x^0_{\mathrm{X}})$ but it was shown in reference \citen{CLO04} that
at least CVM in the tetrahedron-octahedron approximation
has to be used in order to obtain a good agreement between 
classical nucleation theory predictions and kinetic Monte Carlo simulations.
The comparison of the stationary cluster size distributions 
in the solid solution obtained from cluster dynamics simulations 
with the ones predicted by classical nucleation theory leads to the 
same conclusion: one has to use CVM to calculate the nucleation free energy
in order to get the right distribution (\Fig{size_distrib_fig})\footnote{The 
difference between the distribution given by equation \ref{distri_amas_eq} and
the observed stationary distribution during cluster dynamics simulations
for size equal or slightly smaller than the critical one can be attributed
to the Zeldovitch factor. Indeed, for the critical size, there should be a factor 
$1/2$ between the two distributions \cite{MARKOV}.}.
This result can appear quite surprising: although no information concerning
the nucleation free energy has been used to run cluster dynamics simulations, 
this method leads to the stationary size distribution corresponding to 
the CVM. It seems that the lattice gas thermodynamic description used by the cluster
dynamics manages to take into account short range order effects 
which are considered within CVM, in opposition to ideal and regular solid solution
approximations.

\section{Effect of the overlap of diffusion fields}

As seen in the previous section, discrepancies between cluster dynamics 
and kinetic Monte Carlo simulations appear for high supersaturations,
around 1~at.\% at $T=450$°C for instance.
Although the maximal precipitate density and its variations with
temperature and supersaturation are correctly predicted,
the precipitation kinetics as given by cluster dynamics are too slow.  
One possible origin of such a discrepancy 
might be in the expression for the condensation rate $\beta_{\nX}$:
the later is computed solving, under stationary conditions, the diffusion equation
around the cluster isolated in an infinite medium.
This assumption looks reasonable for dilute solutions but 
its validity is not so clear for high supersaturations
when clusters are quite close from each other:
overlap of diffusion fields might alter the rate of solute impingement at clusters.
Therefore, in order to see if a better agreement can be obtained
between cluster dynamics and kinetic Monte Carlo simulations,
it is worth trying to describe kinetic effects 
arising from interactions between the diffusion fields 
created around each precipitate and to incorporate those in 
cluster dynamics simulations. 

In this purpose, we follow the approach proposed by Brailsford \etal \cite{BRA76}
and already used by Smetniansky-De-Grande and Barbu 
in their modeling of the precipitation kinetics in Fe-Cu alloys
\cite{SME94,SME97}.
When calculating the monomer condensation rate on a given cluster,
we assume that at distances greater than $r^{ext}$, 
one half of the mean distance between precipitates, monomers feel 
influence of the cluster through its diffusion field 
as well as the one arising from other clusters through an effective medium
characterized by a parameter $k$.
The rate at which this effective medium absorbs monomers is given by
$D_{\mathrm{X}}k^2(C_1-C_1^{eq})$, $C_1$ and $C_1^{eq}$ being
respectively the instantaneous and the equilibrium monomer concentrations.
This leads to the following expression of the condensation rate, 
depending on whether the radius $r_{\nX}$ of the cluster is smaller or greater than $r^{ext}$
(\cf appendix \ref{condensation_app}),
\begin{subequations}
\begin{alignat}{2}
	\beta_{\nX} &= 4 \pi \frac{D_{\mathrm{X}}}{\Omega}
	r_{\nX} C_1 \frac{1+kr^{ext}}{1+k(r^{ext}-r_{\nX})} &, \quad  r_{\nX} \leq r^{ext},\\
	\beta_{\nX} &= 4 \pi \frac{D_{\mathrm{X}}}{\Omega} r_{\nX} C_1 (1+kr_{\nX}) &, \quad  r_{\nX} \geq r^{ext}.
\end{alignat}
\label{beta_diffusion_eq}
\end{subequations}
Comparing this expression with the previous one (\Eq{eq:beta_n})
where precipitates were assumed to be isolated, we see that
the consideration of interactions between the different cluster diffusion fields
through the introduction of this effective medium leads to a 
greater condensation rate.

So as to ensure that cluster dynamics simulations converge to equilibrium, 
\ie a two-phases system consisting of Al$_3$X precipitates embedded in 
a solid solution of composition $x^{eq}_{\mathrm{X}}$, evaporation rates
need to be multiplied by the same factor as the condensation rates.
They are now given by
\begin{subequations}
\begin{alignat}{2}
	\alpha_{\nX} &= \alpha^0_{\nX} \frac{1+kr^{ext}}{1+k(r^{ext}-r_{\nX-1})} &, \quad  r_{\nX-1} \leq r^{ext},\\
	\alpha_{\nX} &= \alpha^0_{\nX} (1+kr_{\nX-1}) &, \quad  r_{\nX-1} \geq r^{ext},
\end{alignat}
\label{alpha_diffusion_eq}
\end{subequations}
where $\alpha^0_{\nX}$ is the evaporation rate of an isolated cluster
as given by equation \ref{an_eq2}.
Through the parameters $k$ and $r^{ext}$, the evaporation rates
now depend on the solid solution considered.

As the effective medium we introduced
has to be equivalent to the cluster assembly, 
the parameter $k$ is calculated so as to ensure that the quantity
of monomers absorbed by this effective medium
is equal to the quantity consumed by all clusters,
\begin{equation}
D_{\mathrm{X}} k^2 \left(C_1 - C_1^{eq}\right) = 2 J_{1\to2} + \sum_{\nX\geq2}{J_{\nX\to\nX+1}}.
\label{k_self_cons_eq}
\end{equation}
This equation has to be solved self-consistently at each step
of cluster dynamics simulations.
Starting from the value of $k$ corresponding to the previous iteration
or from 0 for the first step,
equations \ref{beta_diffusion_eq} and \ref{alpha_diffusion_eq}
allow to calculate the cluster condensation and evaporation rates
and thus the different fluxes.
Using equation \ref{k_self_cons_eq} a new value of the parameter
$k$ can be computed. To do so, 
the equilibrium monomer concentration $C_1^{eq}$ is required.
It is simply given by the equilibrium cluster size distribution 
\ref{distri_amas_eq} when the nucleation free energy is null,
\begin{equation}
C_1^{eq} = \exp{\left(-(36\pi)^{1/3}\sigma_1 / kT\right)}.
\end{equation}

\begin{figure}[!bp]
\begin{center}
\includegraphics[width=0.8\linewidth]{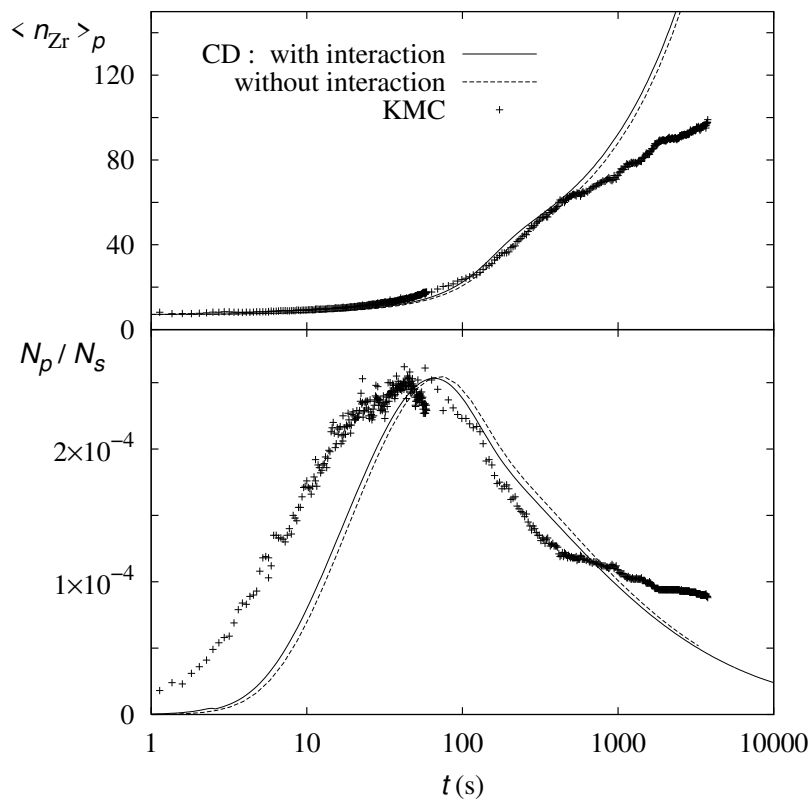}
\end{center}
\caption{Precipitation kinetics of a supersaturated solid solution
of nominal concentration $x^0_{\mathrm{Zr}}=1$~at.\%
at $T=450$°C.
Lines correspond to cluster dynamics simulations with or without
interaction between the diffusion fields created around each cluster
and symbols to kinetic Monte Carlo simulations.}
\label{kinetics_diffusion_field}
\end{figure}

Precipitation kinetics obtained with cluster dynamics when 
considering interactions between the cluster diffusion fields
are not really different from the ones previously obtained 
without such an interaction. At low supersaturations, 
no difference is observed whereas at higher supersaturations 
(\Fig{kinetics_diffusion_field}) this interaction slightly speeds up 
the precipitation kinetics. Nevertheless, this does not really 
improve the agreement with kinetic Monte Carlo simulations 
and kinetics obtained with cluster dynamics still appear too slow.
Therefore the discrepancies appearing at high supersaturations
do not arise from the assumption of isolated clusters used to calculate 
the condensation rate but should have another origin.
We checked too that the use of stationary conditions to compute 
the condensation rate is correct by using the whole solution,
\ie incorporating the transition state, of the diffusion equations \cite{WAI58}
but this does not lead to a better agreement.
Another possible origin of the discrepancy could be the use of 
a single parameter, their size, to describe clusters which involves
considering clusters as spherical particles. This assumption might be too crude
when computing the condensation rate but we did not manage to check
this point as this spherical approximation is needed to solve 
the long-range diffusion equations.

As the correction taking into account the overlap of cluster diffusion 
fields does not really change cluster dynamics simulations, we will 
not consider it in the following section when comparing simulated 
kinetics with experimental data:
the condensation and evaporation rates will be those
of isolated clusters as given by equations \ref{eq:beta_n} and \ref{an_eq2}.

\section{Comparison with experimental data}

Unlike kinetic Monte Carlo which can only simulate rapid processes
(early stages of precipitation, high supersaturations), 
cluster dynamics can be used for low supersaturations and long annealing times.
This allows a direct comparison with experimental studies.
The precipitation kinetics in Al-Zr\cite{IZU69,IZU69b,RYU69,NES72,ZED86,VEC87,ROB01}
as well as in Al-Sc system \cite{DRI84,SAN87,HYL92,JO93,MAR01,NOV01,MAR02,MAR02T,SEI02,ROB03}
has been studied by several groups.
However, for the Al-Zr system, experimental data can hardly be directly
compared to cluster dynamics simulations. Indeed, one observes 
two kinds of precipitates, some spherical ones and some rod-like shaped ones.
Moreover, for low supersaturations, nucleation occurs heterogeneously 
and, Al-Zr being peritectic, high supersaturations are unreachable.
Another difficulty arises from the fact that the L1$_2$ structure is unstable 
and that the stable DO$_{23}$ structure of Al$_3$Zr precipitates appears 
for long enough annealing times.
On the contrary, Al-Sc system appears to be a good candidate to 
compare simulated precipitation kinetics with experimentally
observed ones as experimental data allowing to quantify the homogeneous 
precipitation of the L1$_2$ structure of Al$_3$Sc exists.

\begin{figure}[!bp]
	\begin{center}
		\includegraphics[width=0.8\textwidth]{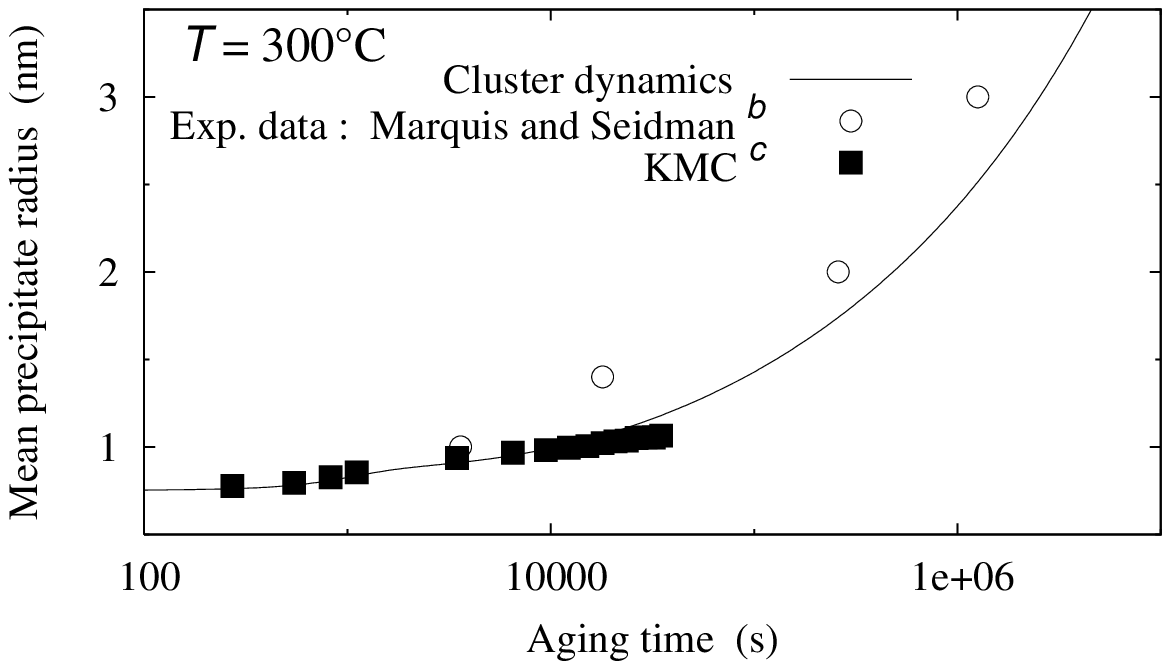}
		\includegraphics[width=0.8\textwidth]{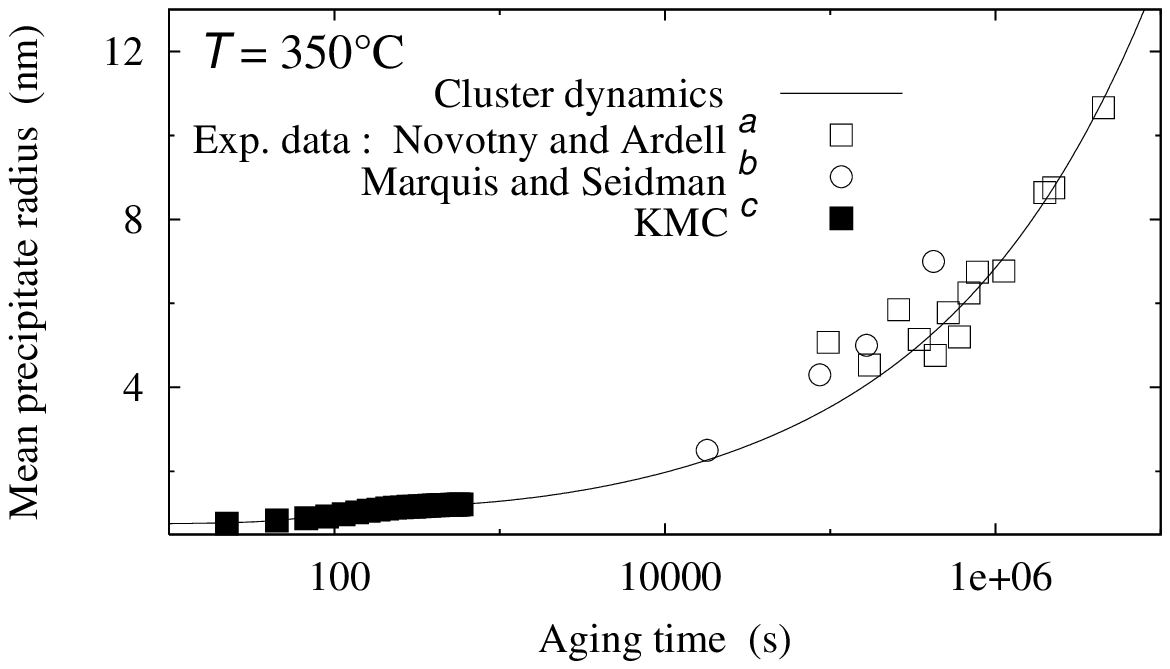}
		\includegraphics[width=0.8\textwidth]{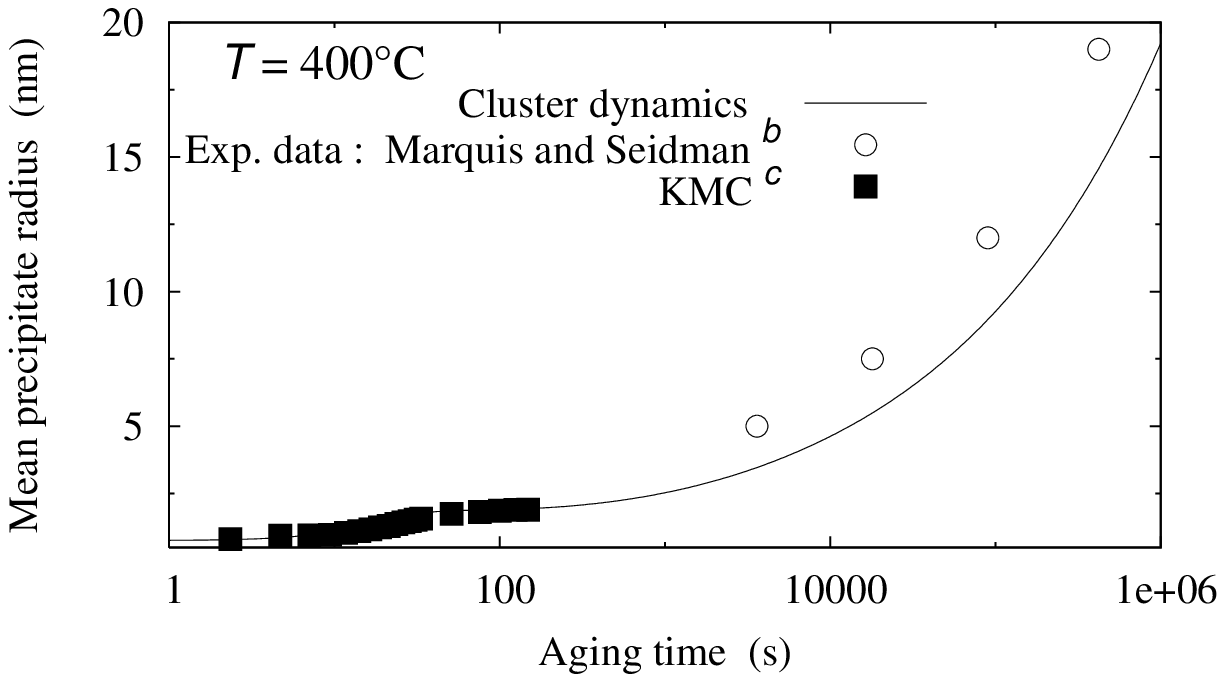}
	\end{center}
	        \hspace{0.1\textwidth}
		$^{a}$~Ref.~\citen{NOV01}, 
		$^{b}$~Ref.~\citen{MAR01,MAR02,MAR02T}, 
		$^{c}$~Ref.~\citen{CLO04}
	\caption{Mean precipitate radius as a function of the aging time
	for a solid solution of composition $x^0_{\mathrm{Sc}}=0.18$~at.\%
	at different temperature ($T=300$, 350, and 400°C) 
	as given by cluster dynamics and compared to experimental
	data\cite{NOV01,MAR01,MAR02,MAR02T} 
	and to kinetic Monte Carlo results\cite{CLO04,CLO04T}.
	The cutoff radius used for cluster dynamics and KMC is 
	$r_{\mathrm{X}}^*\sim0.75$nm ($\nX^*=27$).}
	\label{fig:cinetique1_exp}
\end{figure}

\begin{figure}[!bp]
	\begin{center}
		\subfigure{\includegraphics[width=0.49\textwidth]{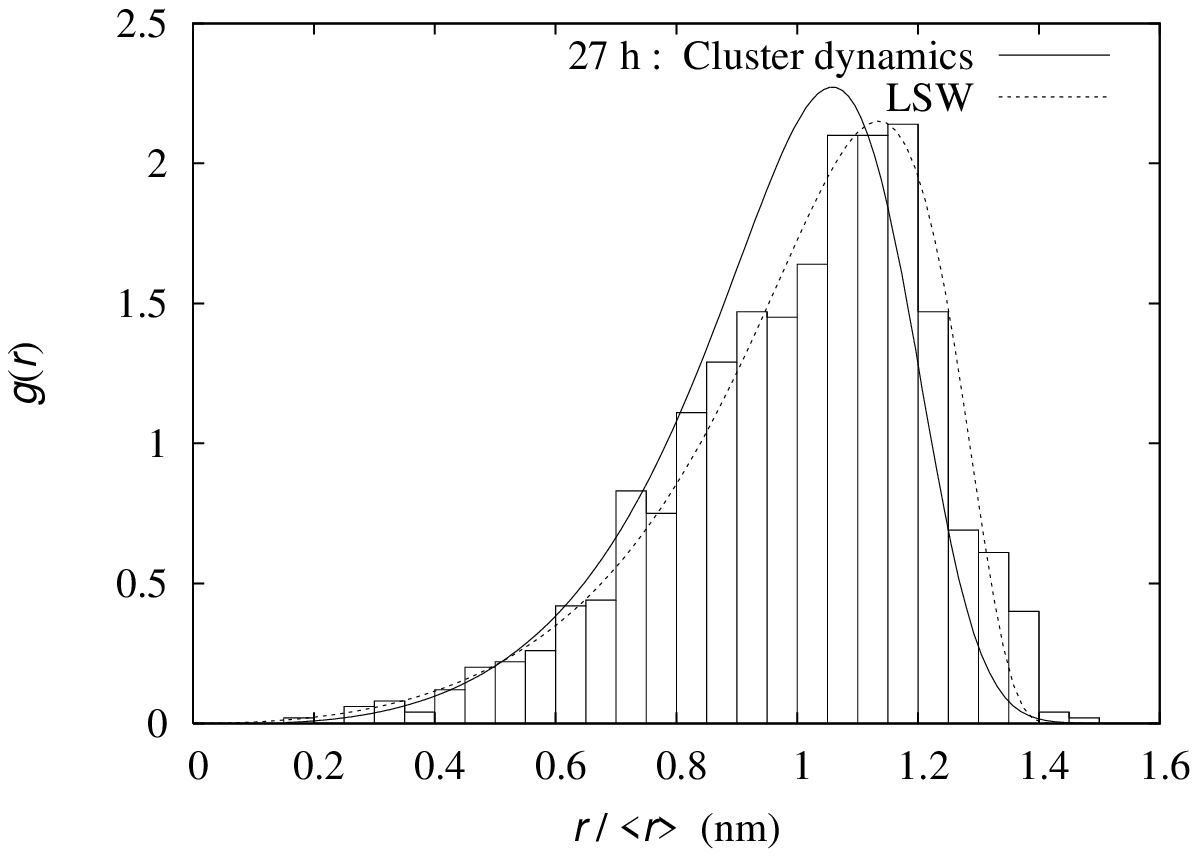}}
		\subfigure{\includegraphics[width=0.49\textwidth]{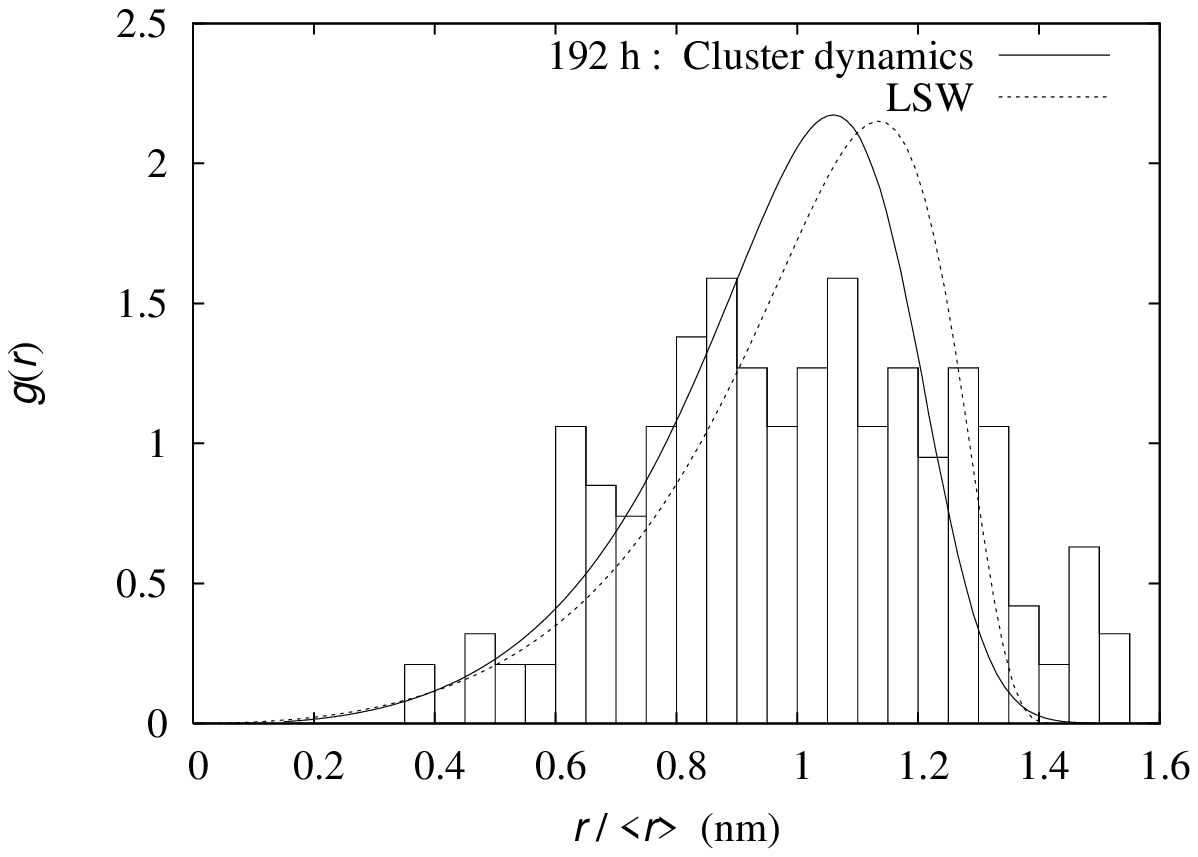}}
		\subfigure{\includegraphics[width=0.49\textwidth]{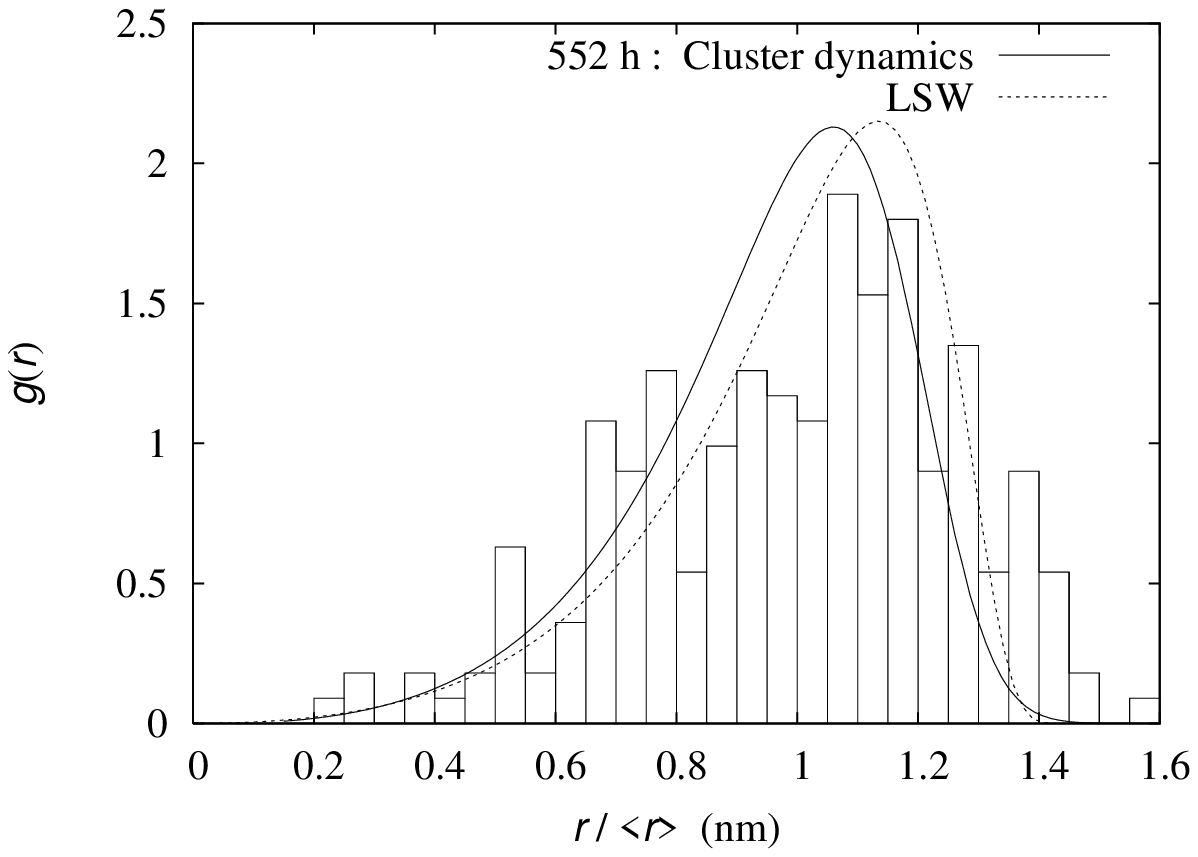}}
		\subfigure{\includegraphics[width=0.49\textwidth]{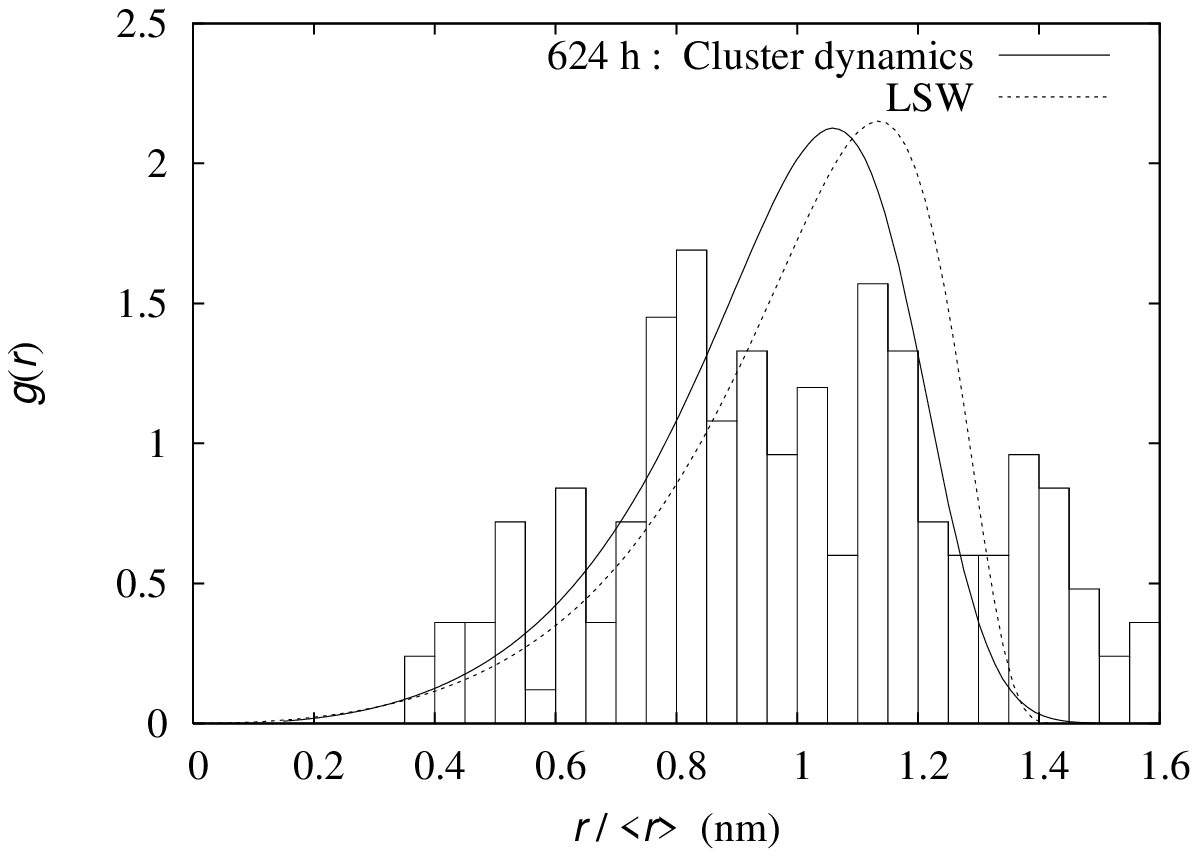}}
		\subfigure{\includegraphics[width=0.49\textwidth]{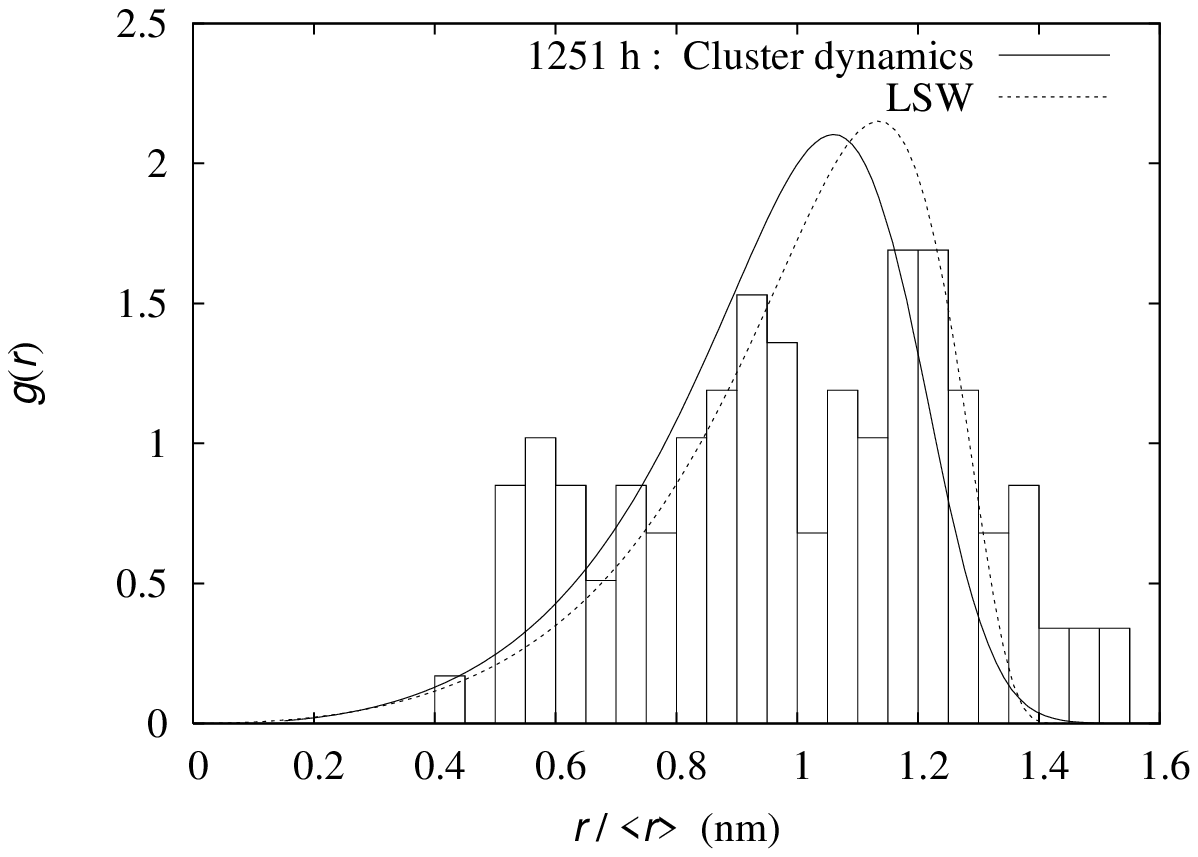}}
		\subfigure{\includegraphics[width=0.49\textwidth]{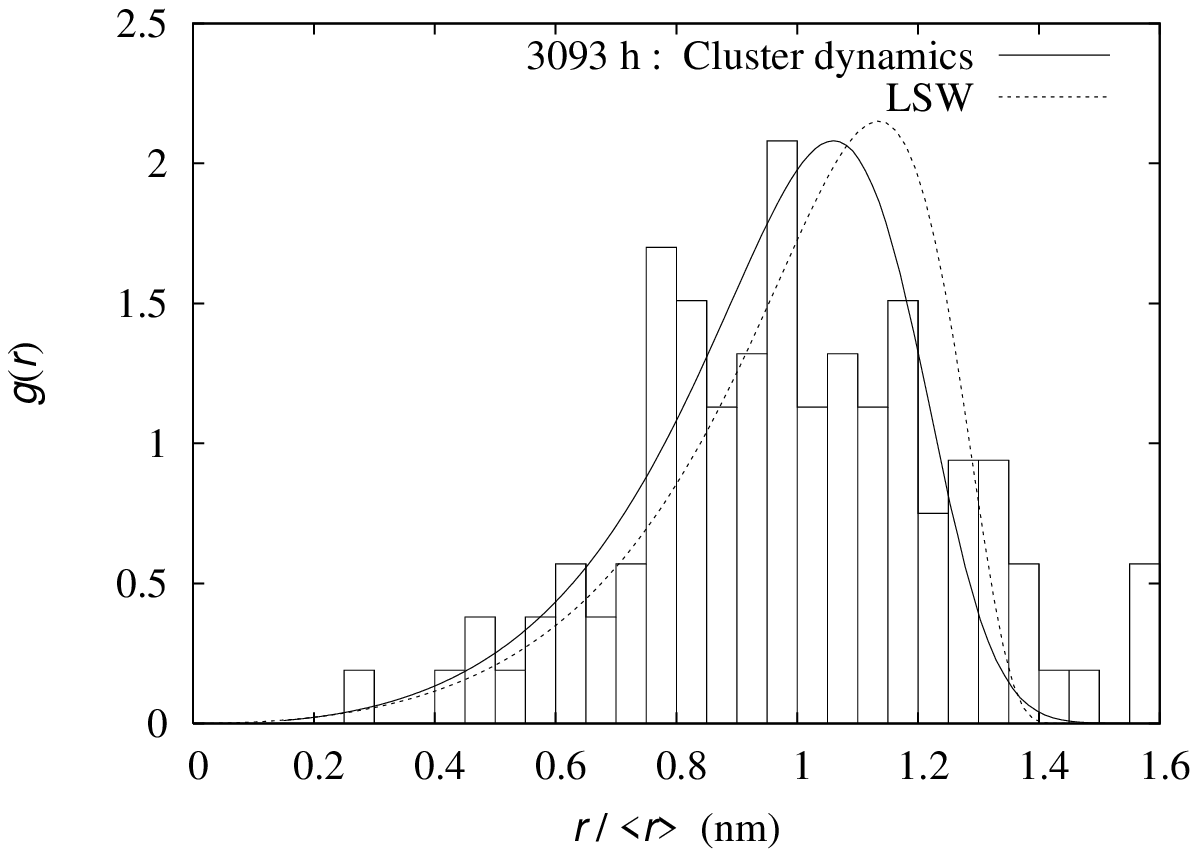}}
	\end{center}
	\caption{Normalized precipitate size distributions $g(r)$ for an Al-0.18~at.\%Sc
	solid solution at $T=350^{\circ}$C obtained with cluster dynamics simulations and compared to experimental
	data \cite{NOV01} as well as to LSW distribution.}
	\label{fig:distri_exp}
\end{figure}

Novotny and Ardell \cite{NOV01} as well as Marquis \etal \cite{MAR01,MAR02,MAR02T} 
observed the coarsening behavior of different Al-Sc alloys. 
For the higher supersaturations studied ($x^0_{\mathrm{Sc}}=0.18$~at.\%), 
the authors concluded that precipitation occurs homogeneously, at least
for temperatures not too high ($T \lesssim 350$°C). 
Therefore, precipitation kinetics of the same alloy can be simulated using
cluster dynamics. 
For the three different temperatures studied ($T=300$, 350, and 400°C),
it appears that the cluster dynamics equation manage to reproduce the variation
with time of the mean precipitate radius (Fig.~\ref{fig:cinetique1_exp}):
the agreement is really good with Novotny's data and reasonable with Marquis' ones.
Notice that the parameters values used in the cluster dynamics were deduced
from the atomic diffusion model used in the kinetic Monte Carlo simulations
of the short annealing times of the same Al-Sc alloy maintained at the same temperatures.
Cluster dynamics, with a single set of parameters, thus reproduces
atomic simulations at short times and gives a safe extrapolation thereof
to the range of annealing times that can be compared with experimental data.
It should be noticed too that cluster dynamics manages to catch the variation with the annealing temperature
of the coarsening kinetics:
the experimental observed speed up of the cluster growth rate 
by a factor $\sim10$ when going from $T=300$ to 400°C
is well predicted.
The good agreement found for the lower temperatures ($T=300$ or 350°C) 
depreciates a little bit at higher temperatures. 
This is due to the fact that, experimentally, for a given nominal concentration one 
reduces the supersaturation by increasing the temperature and thus
favors heterogeneous compared to homogeneous precipitation. 
There are then less precipitates, these ones being larger.

In figure \ref{fig:distri_exp}, we compare
the precipitate size distribution obtained from cluster dynamics
with the experimentally ones measured by Novotny
and Ardell \cite{NOV01}.
For the different annealing times studied the two normalized 
distributions are close. Actually, cluster dynamics leads to 
a normalized size distribution not really different from 
its asymptotic limit, the Lifshitz-Slyozov-Wagner (LSW) distribution. 
Novotny and Ardell already concluded to a good agreement 
between experimental size distributions and the LSW distribution, 
which corresponds to the same agreement found here with the distributions
obtained from cluster dynamics.

Hyland \cite{HYL92}, using TEM, measured the variation with time of the precipitate
density for two different annealing temperatures in an alloy containing
0.11~at.\%~Sc. 
It was not possible to reproduce satisfactorily these experimental data with cluster dynamics.
Indeed the simulated densities are higher by at least one order 
of magnitude than the experimental ones.
The number of precipitates experimentally observed depends on the spatial
resolution of the observation technique. This can be reproduced by imposing
in the simulations a cutoff radius below which the clusters are ignored,
even though they might be supercritical.
The densities calculated from the simulations are really sensitive to the
cutoff radius used to count precipitates. It is possible
to obtain lower densities by increasing this cutoff radius 
but, then, the time scale of the simulations completely differs
from the experimental one.
The origin of this discrepancy remains an open question.

\section{Conclusions}

Precipitation kinetics of Al$_3$Zr and Al$_3$Sc in aluminum supersaturated solid solution
has been modeled using the cluster dynamics technique.
The only input parameters are the solute diffusion coefficients 
and the precipitate / solid solution interface free energies. 
In contrast with other mesoscopic modeling techniques based on Kampmann and Wagner 
approach, cluster dynamics in its strict sense does not require a definition
of the nucleation free energy:
this is not an input parameter of the model but rather
a by-product of the cluster gas thermodynamic underlying the cluster dynamics equations.
We showed that this description leads to a stationary cluster size distribution
in the metastable solid solution corresponding to the one predicted by the classical nucleation 
theory provided CVM is used to calculate the nucleation free energy, in agreement
with previous study.

Cluster dynamics is shown to be very sensitive to the interface free energy: 
the size dependence of the interface free energy, which is revealed 
by the direct computation of the free energy of small clusters,
is crucial in making the cluster dynamics calculations
reproduce kinetic Monte Carlo simulations based on the same
atomistic parameters.
When doing so, the precipitation kinetics obtained agree well with kinetic Monte 
Carlo simulations: the variations of the precipitate density and of their mean
size are well reproduced. 
Nevertheless, for high supersaturations, kinetics modeled by cluster dynamics
appear to be slightly too slow.
We showed that this discrepancy cannot be solved by an improved kinetic model
taking into account the overlap of the cluster diffusion fields.

The cluster dynamics technique as described manages to well reproduce experimental data,
despite the fact that the time scales in real experiments are several order
of magnitude larger than those in kinetic Monte Carlo.
The present study therefore validates, in a quantitative manner, the proposed
multiscale scheme for homogeneous precipitation kinetics:
atomic scale parameters are obtained from available experimental
quantities and ab-initio calculations;
mesoscale parameters are deduced from the latter.
Cluster dynamics as used in this work, \ie in its strict sense,
then appears as the good tool to extrapolate kinetic Monte
Carlo simulations, which reveal full details of the nucleation process,
to a wider range of annealing times, of temperatures and of supersaturations.
It provides an effective method to accomplish quantitative transition of scales,
at least in the low solubility systems which have been considered in this work.

\appendix

\section{Equilibrium cluster size distribution}
\label{distri_app}

In cluster dynamics, the size distribution of the clusters
in a solid solution at equilibrium
is used to deduce the monomer evaporation rates from the condensation rates.
This distribution is needed by classical nucleation theory too so as to 
estimate the cluster size distribution in the metastable solid solution
and to deduce from it the steady-state nucleation rate.
In the following, we show how this distribution can be obtained 
for an undersaturated solid solution, as needed by cluster dynamics, 
the classical nucleation theory extending the validity of this distribution
to supersaturated solid solutions.

The cluster size distribution in a solid solution at equilibrium, 
can be established quite easily as long as
the solid solution is dilute \cite{FRENKEL,KASHCHIEV}.
Considering an assembly composed of $N_{\nX}$ clusters containing
$\nX$ solute atoms and of free energy $G_{\nX}$, the total free energy
of the system can be written
\begin{equation}
	N_s G = \sum_{\nX=1}^{\infty}{N_{\nX} G_{\nX}} + kT \log{\left(\mathcal{W}\right)},
\end{equation}
where $N_s$ is the number of sites that can be occupied by a cluster
and $\mathcal{W}$ is the number of different configurations accessible to
the cluster assembly.
Assuming that each cluster whatever its size lies only on one site
and neglecting around each cluster all excluded sites which cannot be 
occupied by any other cluster, this number is simply given by
\begin{equation}
	\mathcal{W} = \frac{N_s!}{\left(N_s-\sum_{\nX=1}^{\infty}{N_{\nX}}\right)!
	\ \prod_{\nX=1}^{\infty}{N_{\nX}!} }.
\end{equation}
The concentration of each size class being $C_{\nX}=N_{\nX}/N_s$, 
Stirling formula leads to the following estimation for the total free energy,
\begin{equation}
\begin{split}
G =& \sum_{\nX=1}^{\infty}{C_{\nX} G_{\nX}} 
    + kT \sum_{\nX=1}^{\infty}{C_{\nX}\log{(C_{\nX})}} \\
   &+ kT \left(1-\sum_{\nX=1}^{\infty}{C_{\nX}}\right)\log{\left(1-\sum_{\nX=1}^{\infty}{C_{\nX}}\right)}.
\end{split}
\label{G_distri_amas}
\end{equation}
The equilibrium concentrations $\bar{C}_{\nX}$ are obtained by minimizing
this free energy while imposing that the solute concentration equals 
a fixed value $x^0_{\mathrm{X}}$:
\begin{equation}
\sum_{\nX=1}^{\infty}{\nX C_{\nX}} = x^0_{\mathrm{X}}.
\label{bilan_mat_eq}
\end{equation}
This can be done quite easily by considering the grand canonic free energy
\begin{equation}
\mathcal{A} = G + (1-2 \sum_{\nX=1}^{\infty}{\nX C_{\nX}}) \mu,
\end{equation}
where we have introduced a Lagrange multiplier $\mu$ to ensure
the constraint \ref{bilan_mat_eq}. 
$\mu=(\mu_{\mathrm{X}}-\mu_{\mathrm{A}})/2$ is nothing else than 
the effective chemical potential, \ie half the difference between 
the chemical potentials of the solvent A and the solute B.
The minimization of this grand canonic free energy leads to the following solution 
for the cluster size distribution
\begin{equation}
\frac{\bar{C}_{\nX}}{1-\sum_{j=1}^{\infty}{\bar{C}_j}} = \exp{\left(-\frac{G_{\nX}-2\nX\mu}{kT}\right)}.
\label{Cn_distri_amas}
\end{equation}
$\Delta G_{\nX} = G_{\nX}-2\nX\mu$ is the cluster formation free energy 
relative to the solid solution. It can be estimated using the capillary 
approximation.
As we made the assumption of a dilute solid solution to estimate the total free energy $G$,
the sum appearing in the left hand side of equation \ref{Cn_distri_amas} can be neglected,
so the equilibrium cluster size distribution is simply given by
\begin{equation}
\bar{C}_{\nX} = \exp{\left(-\Delta G_{\nX}/kT\right)}.
\end{equation}

\section{Condensation rate}
\label{condensation_app}

Monomer condensation rates on clusters are the main input of cluster dynamics. 
Assuming that the precipitation kinetics is controlled by the solute long
range diffusion, the condensation rates are obtained by solving the diffusion 
equation for the solute in the solvent matrix to a spherical cluster 
with the following boundary conditions:
the solute concentration is zero at the cluster (in order to describe 
solute trapping at the cluster) 
and far from this cluster the monomer concentration is equal to that 
corresponding to cluster dynamics simulation.
Two cases can be distinguished, depending on whether the cluster is assumed to be isolated 
or embedded in an effective medium which simulates the trapping to other clusters,
\ie the effect of the overlap of diffusion fields.

\subsection{Isolated precipitate}

The rate at which monomers condense on an isolated cluster of radius $r_{\nX}$
is obtained by solving the stationary diffusion equation
\begin{equation}
D_{\mathrm{X}} \Delta C_1(r) = 0,
\end{equation}
with the following boundary conditions:
$C_1(r_{\nX})=0$ at the cluster-matrix interface\footnote{Actually, 
at the interface between the cluster and the matrix,
the monomer concentration is equal to the equilibrium one corrected from Gibbs-Thomson effect.
Taking this concentration equal to 0 is equivalent to assume that the cluster does not 
emit any monomer and allows one to directly obtain the condensation rate.}
and $C_1(r\to\infty)=C_1^{\infty}$ far away from the cluster.
The monomer concentration profile around the precipitate is thus
\begin{equation}
C_1(r) = C_1^{\infty} \frac{r-r_{\nX}}{r}.
\end{equation}
The condensation rate is given by the integral over the cluster surface
of the incoming flux of monomers,
\begin{equation}
\begin{split}
	\beta_{\nX} &= - 4 \pi {r_{\nX}}^2 \frac{D_{\mathrm{X}}}{\Omega}
	\left. \frac{\partial{C_1(r)}}{\partial{r}} \right|_{r=r_{\nX}} \\
	&= 4 \pi \frac{D_{\mathrm{X}}}{\Omega} r_{\nX} C_1^{\infty},
\end{split}
\end{equation}
where $\Omega$ is the mean atomic volume corresponding to one lattice site.

\subsection{Precipitate in an absorbing medium}

In order to account for the overlap of the diffusion fields around the many clusters 
in the matrix, one can compute the diffusion flux to one specific cluster
embedded in an effective absorbing medium. The latter extends from infinity
to a distance $r^{ext}$ of the cluster, equal to the mean inter-cluster distance.
This effective medium is characterized by a constant $k$ giving the rate at which monomers are absorbed,
$D_{\mathrm{X}} k^2 \left(C_1(r)-C_1^{eq}\right)$. 
Therefore, for precipitates of radius smaller than $r^{ext}$, the condensation rate is obtained
by solving the set of equations
\begin{subequations}
\begin{align}
D_{\mathrm{X}} \Delta C_1(r) = 0,& \quad r \leq r^{ext} \\
D_{\mathrm{X}} \Delta C_1(r) - D_{\mathrm{X}} k^2 \left(C_1(r)-C_1^{eq}\right) = 0,&  \quad r \geq r^{ext}.
\end{align}
\end{subequations}
The same boundary conditions apply  as previously on the monomer concentration
with the additional condition that the diffusion flux has to be continuous at $r=r^{ext}$. 
The resolution of this diffusion problem gives the following condensation rate,
\begin{equation}
	\beta_{\nX} = 4 \pi \frac{D_{\mathrm{X}}}{\Omega} r_{\nX} C_1^{\infty}
\frac{1+kr^{ext}}{1+k(r^{ext}-r_{\nX})}.
\end{equation}

For precipitates of size $r_n>r^{ext}$ one has simply to solve the equation
\begin{equation}
D_{\mathrm{X}} \Delta C_1(r) - D_{\mathrm{X}} k^2 \left(C_1(r)-C_1^{eq}\right) = 0,
\end{equation}
still with the boundary conditions $C_1(r_{\nX})=0$ and $C_1(r\to\infty)=C_1^{\infty}$.
One obtains for the condensation rate 
\begin{equation}
	\beta_{\nX} = 4 \pi \frac{D_{\mathrm{X}}}{\Omega} r_{\nX} C_1^{\infty}
(1+kr_{\nX}).
\end{equation}

\begin{ack}
    The authors are grateful to Prof. A.~J. Ardell and Prof. D.~N. Seidman
    for providing experimental data.
    They thank too Dr. J.-L. Bocquet, Pr. Y. Bréchet, Dr. J. Dalla Torre,
    Pr. P. Guyot, Dr. B. Legrand, Dr. J. Lépinoux,
    Dr. M. Nastar, Dr. C. Sigli and Dr. F. Soisson for fruitful discussions
    as well as Dr. M.~J. Stowell for his careful reading of the manuscript.
    Part of this work has been funded by the joint research program ``Precipitation''
    between Alcan, Arcelor, CNRS, and CEA.
\end{ack}

\bibliographystyle{elsart-num}
\bibliography{clouet05}

\begin{thebibliography}{10}
\expandafter\ifx\csname url\endcsname\relax
  \def\url#1{\texttt{#1}}\fi
\expandafter\ifx\csname urlprefix\endcsname\relax\def\urlprefix{URL }\fi

\bibitem{ROB01}
J.~D. Robson, P.~B. Prangnell, Dispersoid precipitation and process modelling
  in zirconium containing commercial aluminium alloys, Acta Mater. 49 (2001)
  599--613.

\bibitem{RYU69}
N.~Ryum, Precipitation and recrystallization in an {A}l-0.5~wt.\%~{Z}r alloy,
  Acta Metall. 17 (1969) 269--278.

\bibitem{NES72}
E.~Nes, Precipitation of the metastable cubic {A}l$_3${Z}r-phase in
  subperitectic {A}l-{Z}r alloys, Acta Metall. 20 (1972) 499--506.

\bibitem{HYL92}
R.~W. Hyland, Homogeneous nucleation kinetics of {A}l$_3${S}c in a dilute
  {A}l-{S}c alloy, Metall. Trans. A 23 (1992) 1947--1955.

\bibitem{MAR01}
E.~A. Marquis, D.~N. Seidman, Nanoscale structural evolution of {A}l$_3${S}c
  precipitates in {A}l-{S}c alloys, Acta Mater. 49 (2001) 1909--1919.

\bibitem{NOV01}
G.~M. Novotny, A.~J. Ardell, Precipitation of {A}l$_3${S}c in binary {A}l-{S}c
  alloys, Mater. Sci. Eng. A318 (2001) 144--154.

\bibitem{CLO04}
E.~Clouet, M.~Nastar, C.~Sigli, Nucleation of {Al$_3$Zr} and {Al$_3$Sc} in
  aluminum alloys: from kinetic {M}onte {C}arlo simulations to classical
  theory, Phys. Rev. B 69 (2004) 064109.

\bibitem{CLO04T}
E.~Clouet, Séparation de phase dans les alliages {A}l-{Z}r-{S}c: du saut des
  atomes à la croissance de précipités ordonnés, Thèse de doctorat, \'Ecole
  Centrale Paris, http://tel.ccsd.cnrs.fr/documents/archives0/00/00/59/74
  (2004).

\bibitem{GOL95}
S.~I. Golubov, Y.~N. Osetsky, A.~Serra, A.~V. Barashev, The evolution of copper
  precipitates in binary {F}e-{C}u alloys during ageing and irradiation, J.
  Nucl. Mater. 226 (1995) 252--255.

\bibitem{MAT97}
M.~H. Mathon, A.~Barbu, F.~Dunstetter, F.~Maury, N.~Lorenzelli, C.~H. {de
  Novion}, Experimental study and modelling of copper precipitation under
  irradiation in dilute {F}e{C}u alloys, J. Nucl. Mater. 245 (1997) 224--237.

\bibitem{BAR04}
A.~V. Barashev, S.~I. Golubov, D.~J. Bacon, P.~E.~J. Flewitt, T.~A. Lewis,
  Copper precipitation in {F}e-{C}u alloys under electron and neutron
  irradiation, Acta Mater. 52 (2004) 877--886.

\bibitem{LAE04}
L.~Lae, P.~Guyot, C.~Sigli, Cluster dynamics in {A}l{Z}r and {A}l{S}c alloys,
  in: J.~F. Nie, A.~J. Morton, B.~C. Muddle (Eds.), Proceedings of the 9$^{th}$
  International Conference on Aluminium Alloys, Institute of Materials
  Engineering Australasia Ltd, 2004, pp. 281--286.

\bibitem{MAR04}
G.~Martin, Classical nucleation theory and cluster dynamics: the missing link
  to be published.

\bibitem{HAR02b}
A.~{Hardouin Duparc}, C.~Moingeon, {N. Smetniansky-de-Grande}, A.~Barbu,
  Microstructure modelling of ferritic alloys under high flux 1 {M}e{V}
  electron irradiations, J. Nucl. Mater. 302 (2002) 143--155.

\bibitem{WAI58}
T.~R. Waite, General theory of bimolecular reaction rates in solids and
  liquids, J. Chem. Phys. 28 (1958) 103--106.

\bibitem{MAR78}
G.~Martin, The theories of unmixing kinetics of solid solutions, in: Solid
  State Phase Transformation in Metals and Alloys, Les \'Editions de Physique,
  Orsay, France, 1978, pp. 337--406.

\bibitem{NAS04}
M.~Nastar, E.~Clouet, Mean field theories for the description of diffusion and
  phase transformations controlled by diffusion, Phys. Chem. Chem. Phys. 6
  (2004) 3611--3619.

\bibitem{KAT77}
J.~L. Katz, H.~Wiedersich, Nucleation theory without {M}axwell demons, J.
  Colloid and Interface Science 61 (1977) 351--355.

\bibitem{LAN80}
J.~S. Langer, A.~J. Schwartz, Kinetics of nucleation in near-critical fluids,
  Phys. Rev. A 21 (1980) 948--958.

\bibitem{WAG91}
R.~Wagner, R.~Kampmann, Homogeneous second phase precipitation, in: R.~W. Cahn,
  P.~Haasen, E.~J. Kramer (Eds.), Materials Science and Technology, a
  Comprehensive Treatment, Vol.~5, VCH, Weinheim, 1991, Ch.~4, pp. 213--303.

\bibitem{DES99}
A.~Deschamps, Y.~Br\'echet, Influence of predeformation and ageing of an
  {A}l-{Z}n-{M}g alloy -- ii. modeling of precipitation kinetics and yield
  stress, Acta Mater. 47~(1) (1999) 293--305.

\bibitem{ROB03}
J.~D. Robson, M.~J. Jones, P.~B. Prangnell, Extension of the {N}-model to
  predict competing homogeneous and heterogeneous precipitation in {A}l-{S}c
  alloys, Acta Mater. 51 (2003) 1453--1468.

\bibitem{KIK51}
R.~Kikuchi, A theory of cooperative phenomena, Phys. Rev. 81~(6) (1951)
  988--1003.

\bibitem{SAN78}
J.~M. Sanchez, D.~de~Fontaine, The fcc {I}sing model in the cluster variation
  approximation, Phys. Rev. B 17~(7) (1978) 2926--2936.

\bibitem{LANDOLT26}
H.~Bakker, H.~P. Bonzel, C.~M. Bruff, M.~A. Dayananda, W.~Gust, J.~Horvth,
  I.~Kaur, G.~Kidson, A.~D. LeClaire, H.~Mehrer, G.~Murch, G.~Neumann,
  N.~Stolica, N.~A. Stolwijk, Diffusion in solid metals and alloys, in:
  H.~Mehrer (Ed.), {L}andolt-{B}\"ornstein, New Series, Group {III}, Vol.~26,
  Springer-Verlag, Berlin, 1990.

\bibitem{MAR73}
T.~Marumo, S.~Fujikawa, K.~Hirano, Diffusion of zirconium in aluminum,
  Keikinzoku - J. Jpn. Inst. Light Met. 23 (1973) 17.

\bibitem{FUJ97}
S.~I. Fujikawa, Impurity diffusion of scandium in aluminum, Defect Diff. Forum
  143--147 (1997) 115--120.

\bibitem{GOL00}
S.~I. Golubov, A.~Serra, Y.~N. Osetsky, A.~V. Barashev, On the validity of the
  cluster model to describe the evolution of {C}u precipitates in {F}e-{C}u
  alloys, J. Nucl. Mater. 277 (2000) 113--115.

\bibitem{PER84}
A.~Perini, G.~Jacucci, G.~Martin, Cluster free energy in the simple-cubic
  {I}sing model, Phys. Rev. B 29 (1984) 2689--2697.

\bibitem{TOL49}
R.~C. Tolman, The effect of droplet size on surface tension, J. Chem. Phys. 17
  (1949) 333--337.

\bibitem{BUF55}
F.~P. Buff, Spherical interface. {II}. {M}olecular theory, J. Chem. Phys. 23
  (1955) 419.

\bibitem{LEP04}
J.~L{\'e}pinoux, Private communication (2004).

\bibitem{MARKOV}
I.~V. Markov, Crystal Growth for Beginners, World Scientific, Singapore, 1995.

\bibitem{BRA76}
A.~D. Brailsford, R.~Bullough, M.~R. Hayns, Point defect sink strengths and
  void-swelling, J. Nucl. Mater. 60 (1976) 246--256.

\bibitem{SME94}
N.~Smetniansky-De-Grande, A.~Barbu, Study of {C}u precipitation mechanisms in
  {F}e-{C}u 1.34\% at. alloy under electron irradiation, Radiat. Eff. Defects
  Solids 132 (1994) 157--167.

\bibitem{SME97}
N.~Smetniansky-De-Grande, A.~Barbu, Study of {C}u precipitation in {F}e-{C}u
  binary alloys under irradiation. part {II}, Radiat. Eff. Defects Solids 140
  (1997) 313--321.

\bibitem{IZU69}
O.~Izumi, D.~Oelschl{\"a}gel, On the decomposition of a highly supersaturated
  {A}l-{Z}r solid solution, Scripta Metall. 3 (1969) 619--621.

\bibitem{IZU69b}
O.~Izumi, D.~Oelschl{\"a}gel, Structural investigation of precipitation in an
  aluminium alloy containing 1.1 weight per cent zirconium, Z. Metalkd. 60
  (1969) 845--851.

\bibitem{ZED86}
M.~S. Zedalis, M.~E. Fine, Precipitation and {O}stwald ripening in dilute
  {A}l-based-{Z}r-{V} alloys, Metall. Trans. A 17 (1986) 2187--2198.

\bibitem{VEC87}
K.~S. Vecchio, D.~B. Williams, Convergent beam electron diffraction study of
  {A}l$_3${Z}r in {A}l--{Z}r and {A}l--{L}i--{Z}r alloys, Acta Metall. 35
  (1987) 2959--2970.

\bibitem{DRI84}
M.~E. Drits, J.~Dutkiewicz, L.~S. Toropova, J.~Salawa, The effect of solution
  treatment on the ageing processes of {A}l-{S}c alloys, Crystal Res. Technol.
  19 (1984) 1325--1330.

\bibitem{SAN87}
N.~Sano, Y.~Hasegawa, K.~Hono, H.~Jo, K.~Hirano, H.~W. Pickering, T.~Sakurai,
  Precipitation process of {A}l-{S}c alloys, J. Phys. (Paris) C6 (1987)
  337--342.

\bibitem{JO93}
H.-H. Jo, S.-I. Fujikawa, Kinetics of precipitation in {A}l-{S}c alloys and low
  temperature solid solubility of scandium in aluminium studied by electrical
  resistivity measurements, Mater. Sci. Eng. A171 (1993) 151--161.

\bibitem{MAR02}
E.~A. Marquis, D.~N. Seidman, D.~C. Dunand, Creep of precipitation-strengthened
  {A}l({S}c) alloys, in: R.~S. Mishra, J.~C. Earthman, S.~V. Raj (Eds.), Creep
  Deformation: Fundamentals and Applications, TMS, 2002, p. 299.

\bibitem{MAR02T}
E.~A. Marquis, Microstructural evolution and strengthening mechanism in
  {A}l-{S}c and {A}l-{M}g-{S}c alloys, Ph.D. thesis, Northwestern University,
  Evanston, Illinois (2002).

\bibitem{SEI02}
D.~N. Seidman, E.~A. Marquis, D.~C. Dunand, Precipitation strengthening at
  ambient and elevated temperatures of heat-treatable {A}l({S}c) alloys, Acta
  Mater. 50 (2002) 4021--4035.

\bibitem{FRENKEL}
J.~Frenkel, Kinetic Theory of Liquids, Dover Publications, New York, 1955.

\bibitem{KASHCHIEV}
D.~Kashchiev, Nucleation : basic theory with applications, Butterworth
  Heinemann, Oxford, 2000.

\end{thebibliography}

\end{document}